\documentclass[aps,prd,twocolumn,nobalancelastpage,nofootinbib,floatfix]{revtex4-1}

\usepackage{graphicx}
\usepackage{amsmath}
\usepackage{amssymb}
\usepackage{amsfonts}
\usepackage{textcomp}
\usepackage{siunitx}

\begin{document}


\title{Deep inelastic vortex scattering:\\ A third outcome for head-on collisions}


\author{Ethan P. Honda}
\email[]{ehonda@alum.mit.edu}
\affiliation{Melbourne, Florida, 32940, USA}


\date{\today}

\begin{abstract}
Results are presented from numerical simulations of the flat-space nonlinear Maxwell-Klein-Gordon 
equations demonstrating deep inelastic scattering of $m=1$ vortices for a range of Ginzburg-Landau (or Abelian-Higgs)
parameters ($\kappa$), 
impact parameters ($b$), 
and initial velocities ($v_0$).
The threshold ($v_0^*$) of right-angle scattering is explored for head-on ($b=0$) 
collisions by varying $v_0$. 
Solutions obey time-scaling laws, $T\propto \alpha\ln(v_0-v_0^*) $, with $\kappa$-dependent scaling exponents, $\alpha$,
and have $v_0^*$ that appear not to have the previously reported upper bound.
The arbitrarily long-lived static intermediate attractor at criticality ($v_0=v_0^*$) 
is observed to be the $\kappa$-specific $m=2$ vortex solution.
Scattering angles are  observed for off-axis ($b\neq 0$) collisions for a wide range of $b$, $v_0$, and  $\kappa$.
 It is shown that for arbitrarily small impact parameters ($b\rightarrow 0$), the unstable 
 $\kappa$-dependent 
 $m=2$ ``critical" vortex is an intermediate attractor and decays with a  $\kappa$-\emph{independent} 
 scattering angle of $135^{\circ}$, as opposed to either of
 the well-known values of $180^{\circ}$ or $90^{\circ}$ for $b=0$.

\end{abstract}


\maketitle

\section{Introduction}

Topological defects are known to appear in  a wide variety of physical models that involve spontaneous symmetry breaking
of gauge field theories.
In three spatial dimensions, they can appear in the form of  monopoles, strings/vortices, or domain walls
and are often lumped together with their slightly more exotic cousins, textures \cite{PhysRevD.50.2806}.
This paper discusses vortices that are solutions to the axisymmetric Maxwell-Klein-Gordon equations of motion.
They have a conserved topological charge and are 
characterized by a ``winding number" that counts the number 
of times the phase of the complex scalar field  rotates when propagated along a closed loop around the vortex axis.
Such vortices are present in a  range of contexts, including the Abelian-Higgs model of particle physics and  
the Ginzburg-Landau model of condensed matter physics.
While these
models have very different physical interpretations, they reduce to the same dimensionless action and  have
the same dimensionless Maxwell-Klein-Gordon equations of motion, thus  giving rise to common phenomenology of great interest
across many parts of the physics community.

Nielsen and Olesen  explored vortices  in the Abelian-Higgs model 
to better understand the possible string nature of fundamental particles, noting the approximate 
correspondence between a vortex and a Nambu string \cite{NIELSEN197345}.
%
%
A great number of authors have studied
Abelian-Higgs vortices in the context of early universe cosmology, 
where they can appear as  topological defects (cosmic strings) 
 when a  grand unified theory 
 has undergone spontaneous symmetry breaking \cite{Kibble_1976}.
 %
 %
While  not yet experimentally observed, cosmic strings are believed by many to have been present in the early universe, where 
they may have influenced the large-scale structure currently observed in the universe 
\cite{PhysRevD.97.102002,PhysRevD.99.104028,PhysRevD.100.023535}.
%
%
%

%
Abrikosov vortices, on the other hand, are solutions to the Ginzburg-Landau equations of motion that 
describe magnetic flux tubes within 
superconductors. They were
first predicted by Abrikosov in 1957 \cite{Abrikosov_OrigVortex,RevModPhys.76.975} and later observed in the laboratory 
 by Cribier et al. by means of neutron diffraction 
in 1964 \cite{CRIBIER1964106}.
%
This paper adopts the conventions typically used in the context of Ginzburg-Landau vortices, where 
vortices and the ``bulk" supporting them are characterized by a number of parameters.   
The Ginzburg-Landau parameter is given by 
 $\kappa = \Lambda/\xi$
 and is the ratio of the London penetration depth of the magnetic field ($\Lambda$) 
to the coherence length  of the complex scalar field  ($\xi$).
Superconductors with $\kappa < 1/\sqrt{2}$ are referred to as Type-I superconductors; vortices in 
Type-I superconductors attract one another and typically coalesce, breaking down the superconductivity when in abundance.
Superconductors with $\kappa > 1/\sqrt{2}$ are referred to as Type-II superconductors; vortices in 
Type-II superconductors repel one another (do not coalesce) and form flux tube (Abrikosov) lattices when in abundance.
Superconductors with $\kappa = 1/\sqrt{2}$ are  referred to as having critical coupling, and they
support vortices that are noninteracting.%

As topological solitons, vortices have a conserved topological charge and are characterized by their winding number, $m$.   
The equations of motion do not allow an $m=1$ vortex to ``unwind" itself, which ensures stability independent 
of $\kappa$.
However, the stability of vortices with $m>1$ is a $\kappa$-dependent phenomenon.
Bogomol'nyi demonstrated using a trial function approach that  
vortices with $\kappa \leq 1/\sqrt{2}$ can be stable for $m>1$, 
while vortices with  $\kappa>1/\sqrt{2}$ are unstable for  $m>1$
\cite{osti_7309001}.  
The repulsive interaction and instability of vortices with $m>1$ in a Type-II bulk gives rise to very interesting
and nontrivial scattering.

In 1988, it was numerically 
demonstrated  that the head-on collisions of two vortices could lead to deep inelastic 
{right-angle} scattering 
\cite{MORIARTY1988411,SHELLARD1988262}.
This remarkable phenomenon was predicted for a number of models for the noninteracting (critical coupling) cases where one
can obtain Bogomol'nyi equations that reduce second-order partial differential equations (PDEs) 
to simpler, first-order PDEs \cite{RUBACK1988669}.  
While this approach had the benefit of greatly reducing computational demands at a time when such calculations were prohibitive,
the results were only relevant at or near critical coupling.
The noninteracting ($\kappa=1/\sqrt{2}$)
vortices would scatter at right angles, independent of velocity, while repulsive ($\kappa>1/\sqrt{2}$)
vortices would transition
from direct backscattering to right-angle scattering at a $\kappa$-dependent velocity.
In 1992, Myers et al. performed a 
numerical study \cite{PhysRevD.45.1355}
of  vortex scattering,
clearly demonstrating  right-angle scattering
for repulsive vortices with coupling as strong as $\kappa = 2\sqrt{2}$ ($\lambda_{\rm MRS}=32$ in 
their Abelian-Higgs model).
While the authors 
pushed the limits of computation at the time, the breadth of $(b,v_0)$ parameter space,
the resolution of simulations,  
and the range of $\kappa$ were limited.  

This paper discusses new results and enhances previous results of vortex-vortex scattering within this model.  
The previously understood bound on the critical velocity, $v_0^*(\kappa)$, that marks the transition from backscattering to 
right-angle scattering is shown not to exist. 
Further, it is shown that there exist intermediate attractor solutions that are arbitrarily long-lived and demonstrate
a third type of scattering in the limit where the impact parameter goes to zero.
While the results are applicable to both the Abelian-Higgs or Ginzburg-Landau models, conventions are 
adopted that are 
most consistent with the Ginzburg-Landau model.

The remainder of the paper is organized as follows. 
In Sec. \ref{sec:Formalism}, the formalism is defined and the fully general equations of motion are presented.
In Sec. \ref{sec:InitialDataEOM}, the initial data and evolution equations  are discussed.
 Section \ref{sec:HeadOn} discusses head-on collisions and observed scaling laws and shows that in the center of mass
frame,  a static $m=2$ vortex is the intermediate attractor for the threshold of backscattering and right-angle scattering.
Section \ref{sec:OffAxis} presents results from off-axis collisions 
and discusses a new
$\kappa$-independent scattering phenomenon.
Appendix \ref{app:DimensionLessLags} shows how to obtain the dimensionless Maxwell-Klein-Gordon action from 
 the dimensionful Ginzburg-Landau action
 and shows how to relate that dimensionless action to the actions in \cite{SHELLARD1988262} and \cite{PhysRevD.45.1355}.
Appendix \ref{app:ComputationalMethods} discusses the computational methods used.

\section{General Formalism, definitions, and conventions \label{sec:Formalism}}

The dimensionless Maxwell-Klein-Gordon Lagrangian being studied here is easily obtained  
from the Abelian-Higgs or Ginzburg-Landau models by transforming to dimensionless variables
(Appendix \ref{app:DimensionLessLags}), 
\begin{eqnarray}
{L}
&=&  
-\frac{1}{4}{F}^{\mu\nu}{F}_{\mu\nu}  
-\frac{1}{2}g^{\mu\nu}{D}_\mu{\phi} 
                                                      \left( {D}_\nu{\phi}\right)^* 
+  \frac{1}{2}  {\phi}^2 - \frac{1}{4}  {\phi}^4,  \nonumber \\
                                                \label{eqn:DimensionlessBECLagrangian}
\end{eqnarray}
where the gauge covariant derivative and the Maxwell field tensor are given by
\begin{eqnarray}
{D}_\mu{\phi} &=& \left( \frac{1}{\kappa}{\partial}_\mu  - i {A}_\mu\right){\phi}  {\rm \ \ \ and}\\
{F}_{\mu\nu} &=& {\partial}_\mu {A}_\nu - {\partial}_\nu {A}_\mu,
\end{eqnarray}
respectively.
The Maxwell equations are given by
%
\begin{eqnarray}
\frac{1}{\sqrt{-g}} {\partial}_\rho \left( \sqrt{-g} {{F}}^{\sigma\rho} \right) &=& {J}^\sigma {\rm \ \ \ and}  \label{eqn:Max1} \\
{\partial}_{\left[ \alpha \right.} {{F}}_{\left. \mu \nu \right]} &=& 0 \label{eqn:Max2},
\end{eqnarray}
where
\begin{eqnarray}
{J}^\sigma   
 &=& 
- \frac{1}{\kappa}  g^{\sigma\nu}\left( {\phi}_2{\partial}_\nu {\phi}_1 - {\phi}_1 {\partial}_\nu {\phi}_2 \right)
-  \left({\phi}_1^2 + {\phi}_2^2 \right) {{A}}^\sigma 
\hspace{5mm}
\end{eqnarray}
%
is the Maxwell four-current.
The Klein-Gordon equations for the components of the complex scalar field  are given by
%
\begin{eqnarray}
\displaystyle
\Box {\phi}_1
&=&
\displaystyle
- 2 \kappa {{A}}^\sigma {\partial}_\sigma{\phi}_2 
- \kappa {\phi}_2 {\nabla}_\sigma {{A}}^\sigma  
+\kappa^2 {\phi}_1 {{A}}_\sigma {{A}}^\sigma 
\nonumber \\
&&
-  \kappa^2  {\phi}_1 + \kappa^2 {\phi}_1\left( {\phi}_1^2 + {\phi}_2^2\right)
\label{eqn:KG1} {\rm \ \ \ and} \\
\displaystyle
\Box {\phi}_2
&=&
\displaystyle
2 \kappa {{A}}^\sigma {\partial}_\sigma{\phi}_1
+ \kappa {\phi}_1 {\nabla}_\sigma {{A}}^\sigma 
+\kappa^2 {\phi}_2 {{A}}_\sigma {{A}}^\sigma 
\nonumber \\
&&
-  \kappa^2  {\phi}_2 + \kappa^2 {\phi}_2\left( {\phi}_1^2 + {\phi}_2^2\right)
\label{eqn:KG2}, 
\end{eqnarray}
where
\begin{equation}
\Box {\phi}_i
= \frac{1}{\sqrt{-g}} {\partial}_\rho \left( \sqrt{-g} g^{\rho\sigma} {\partial}_\sigma {\phi}_i\right)
\end{equation}
is the covariant d'Alembertian. 
Finally, it is worth noting that 
the signature of the metric is $({-} {+} {+} {+})$ and
all  quantities represent dimensionless variables.

\section{Initial Data and Equations of Motion \label{sec:InitialDataEOM}}

Any analysis of vortex scattering begins with a single static vortex solution.  
The ansatz for an axisymmetric static vortex with winding number $m$ is given by
\begin{equation}
{\phi}_m(R,\theta) = {\phi}(R) e^{ i m\theta}.
\end{equation}
This work discusses scattering of $m=1$ vortices with $\kappa > 1/\sqrt{2}$ and only considers solutions 
with zero charge (${A}_t={E}_R=0$) at $t=0$.
Assuming the Lorentz gauge, 
Eqs. (\ref{eqn:Max1}),
(\ref{eqn:Max2}), (\ref{eqn:KG1}), and (\ref{eqn:KG2})
in cylindrical symmetry
reduce to the following equations:
\begin{eqnarray}
 \partial_R  {\chi}
&=& 
-  2\kappa {A}_\theta \frac{{\phi}}{R}
+ \kappa^2{A}_\theta^2{\phi} 
-\kappa^2 {\phi}
+\kappa^2 \phi^3, 
\hspace{5mm}\\ 
 \partial_{R^2}\left(R{\phi}\right)  &=& \frac{1}{2}{\chi},\\
\partial_R {B}_z %
&=& -\frac{1}{\kappa}  \frac{{\phi}^2}{R}  +  {A}_\theta  {\phi}^2, {\rm \ \ \ and} %
\\
\partial_{R^2}\left( R {A}_\theta \right) &=& \frac{1}{2}{B}_z.	
\end{eqnarray}
The inner ($R=0$) boundary conditions are given by
\begin{eqnarray}
{\chi}(0) 	&=& {\chi}_0, \\
{\phi}(0) &=& 0, \\
{B}_z(0) &=&  \left({B}_z\right)_0, {\rm \ and}\\
{A}_\theta(0) &=&  0,
\end{eqnarray}
since regularity at $R=0$ implies that ${\phi}(R) \approx c R^m$ for
$R\approx 0$.
The initial conditions ${\chi}_0$ and $ ({B}_z)_0$ are varied while ``shooting" on the large-$R$ 
boundary conditions,
\begin{eqnarray}
{\phi}(\infty) &=& 1 {\rm \ \ \ and} \\
{B}_z(\infty) &=& 0.
\end{eqnarray}
As is typical for shooting methods, the fields diverge before reaching the outer boundary, so 
solutions were smoothed and matched to their known large-$R$ behavior.

To numerically simulate a collision of two vortices,  individual static vortices must be boosted and combined.
The vortices are assumed to be positioned symmetrically about the origin and boosted 
at one another in the $x$ direction with equal magnitude velocities, $v_0$, and with
a given impact parameter.
The Maxwell fields of the individually boosted vortices are   added together, 
%
%
%
while the complex scalar fields are combined by multiplying
the magnitudes and adding the phases of the individually boosted vortex solutions.
The ${\Pi}_i = {\partial}_t{\phi}_i$  are obtained by computing the time derivative of the Lorentz-boosted 
static-vortex scalar field solutions in the boosted frame. 
%
%
These initial data are then time evolved with the hyperbolic equations of motion 
in Cartesian coordinates. 
The equations of motion for the electric and magnetic fields 
are obtained from  (\ref{eqn:Max1}) and (\ref{eqn:Max2}), giving
\begin{eqnarray}
{\partial}_t {E}_x &=& 
 {\partial}_y{B}_z 
+ \frac{1}{\kappa}\left( 
{\phi}_2{\partial}_x{\phi}_1 -  {\phi}_1{\partial}_x{\phi}_2  \right) \nonumber \\
&& + {A}_x \left( {\phi}_1^2 + {\phi}_2^2 \right), \\ 
{\partial}_t {E}_y &=& 
 - {\partial}_x {B}_z 
+ \frac{1}{\kappa}\left( 
{\phi}_2{\partial}_y{\phi}_1 -  {\phi}_1{\partial}_y{\phi}_2  \right) \nonumber \\
&& + {A}_y \left( {\phi}_1^2 + {\phi}_2^2 \right), \\
{\partial}_t {E}_z &=& \left( {\partial}_x{B}_y - {\partial}_y {B}_x \right)
+ {A}_z \left( {\phi}_1^2 + {\phi}_2^2 \right), \\ 
{\partial}_t {B}_x &=& 
- {\partial}_y {E}_z,   \\
{\partial}_t {B}_y &=& 
 {\partial}_x {E}_z , {\rm \ \ \ and} \\
{\partial}_t {B}_z &=& 
-\left(  {\partial}_x {E}_y  -   {\partial}_y {E}_x \right). 
\end{eqnarray}
The evolution equations for the components of the gauge potential are obtained from the definition of the Maxwell tensor, 
\begin{eqnarray}
{\partial}_t {A}_x &=& {\partial}_x{A}_t - {E}_x, \\
{\partial}_t {A}_y &=& {\partial}_y{A}_t - {E}_y,  \\
{\partial}_t {A}_z &=& - {E}_z, 
\end{eqnarray}
and the Lorentz gauge condition,
\begin{eqnarray}
{\partial}_t {A}_t &=& {\partial}_x {A}_x + {\partial}_y{A}_y.
\end{eqnarray}
The complex scalar field evolution equations are obtained from (\ref{eqn:KG1}) and (\ref{eqn:KG2}), giving
\begin{eqnarray}
{\partial}_t{\Pi}_1 
&=& {\partial}_x{\Phi}_{1x}  + {\partial}_y{\Phi}_{1y}  
+ 2\kappa\left( -{A}_t{\Pi}_2 +   {A}_x{\Phi}_{2x} + {A}_y{\Phi}_{2y} 
\right)
\nonumber \\
&& - \kappa^2 {\phi}_1 \left( -{A}_t^2 + {A}_x^2 + {A}_y^2 + {A}_z^2 \right)  \nonumber \\
&& + \kappa^2 {\phi}_1 - \kappa^2 {\phi}_1\left( {\phi}_1^2 + {\phi}_2^2 \right)  {\rm \ \ and} \\
{\partial}_t{\Pi}_2
&=& {\partial}_x{\Phi}_{2x}  + {\partial}_y{\Phi}_{2y}  
- 2\kappa\left(-{A}_t{\Pi}_1 +  {A}_x{\Phi}_{1x} + {A}_y{\Phi}_{1y} 
\right)
\nonumber \\
&& - \kappa^2 {\phi}_2 \left( -{A}_t^2 + {A}_x^2 + {A}_y^2 + {A}_z^2 \right)  \nonumber \\
&& + \kappa^2  {\phi}_2 - \kappa^2 {\phi}_2\left( {\phi}_1^2 + {\phi}_2^2 \right), 
\end{eqnarray}
while the definition 
of ${\Pi}_i$ 
and commutation of partial derivatives give
\begin{eqnarray}
{\partial}_t {\Phi}_{1x} &=&  {\partial}_x {\Pi}_1,  \\
{\partial}_t {\Phi}_{1y} &=&  {\partial}_y {\Pi}_1,  \\
{\partial}_t {\Phi}_{2x} &=&  {\partial}_x {\Pi}_2,  \\
{\partial}_t {\Phi}_{2y} &=&  {\partial}_y {\Pi}_2,  \\
{\partial}_t {\phi}_{1} &=& {\Pi}_1, {\rm \ and} \\
{\partial}_t {\phi}_{2} &=& {\Pi}_2. 
\end{eqnarray}
%
%
%
For all  simulations discussed in this paper, these data are evolved with vortices starting 
at positions $(-x_0,y_0)$ and $(x_0,-y_0)$ with velocities $(v_0,0)$ and $(-v_0,0)$, respectively. 
The impact parameter $b = 2 y_0$ is defined to be the distance between the vortices perpendicular to their collision velocity.

\section{Critical head-on scattering \label{sec:HeadOn}}

This section builds upon, and in one case corrects, the current understanding of
head-on ($b=0$) scattering of repulsive 
vortices.
It was clearly shown by \cite{PhysRevD.45.1355} that repulsive vortices
backscatter, $\theta_s=180^\circ$, for low-velocity collisions, $v_0 < v_0^*$,
and right-angle scatter, $\theta_s=90^\circ$, for high-velocity collisions, $v_0 > v_0^*$,
for some  critical velocity, $v_0^*$. 
However, based on the simulations performed at the time, 
it appeared and was conjectured that the critical velocity approached an asymptotic value of 
$v_0 \approx 0.56$ for large $\kappa$.
Those investigations used couplings of $1 \leq \lambda_{\rm MRS} \leq 32$ in the 
Abelian-Higgs model\footnote{See Appendix \ref{app:DimensionLessLags} for detailed model comparison.}, which corresponds to 
$1/2\leq\kappa\leq 2\sqrt{2}$ in the Ginzburg-Landau model,
and used a  coherence length to the lattice spacing ratio ($\xi/{\rm dx}$) of 5:1 at $\kappa=2\sqrt{2}$. 
This work uses a $\xi/{\rm dx}$ ratio of greater than 10:1 and maintains this ratio for increasing $\kappa$ (decreasing $\xi$)
while keeping the domain large enough to also capture the magnetic field dynamics that remain on the length scale 
$\Lambda \approx 1$.  
This approach maintains energy conservation through the life of the simulations to a few parts in  $10^3$ for high 
values of $\kappa$ and a few parts in $10^4$ for smaller $\kappa$ (Appendix \ref{app:ComputationalMethods}).

%
\begin{figure}[t]
\includegraphics[width=0.95\linewidth]{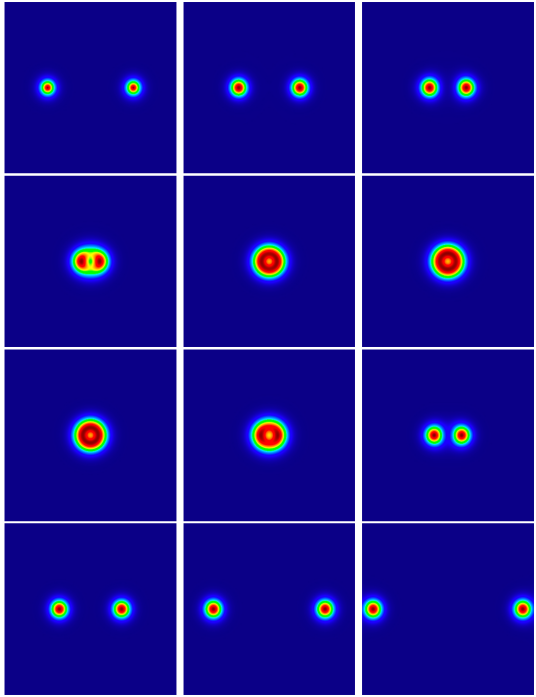}

\caption{ 
Time slices of the energy density for the head-on ($b=0$) 
collision of two $m=1$ vortices with subcritical velocities for $\kappa=1$. 
The plots represent times $t=0, 5, 10, 16.75, 25, 50, 75, 91.75, 100, 107.5, 117.5,\ \rm{and} \ 125$,
displayed from top left to bottom right, respectively.
The simulations were conducted on a $401\times 401$ grid with domain spanning 
$\{x:-10<x<10\}$ and 
$\{y:-10<y<10\}$.
The initial velocities are tuned to within one part in $10^{13}$ of the critical solution,
and the long-lived $m=2$ static intermediate attractor can be observed ($25\lesssim t \lesssim 91.75$).
The solution demonstrates exact backscatter, $\theta_s = 180^{\circ}$.  
\label{fig:reflect}}
\end{figure}
\begin{figure}[t]
\includegraphics[width=0.95\linewidth]{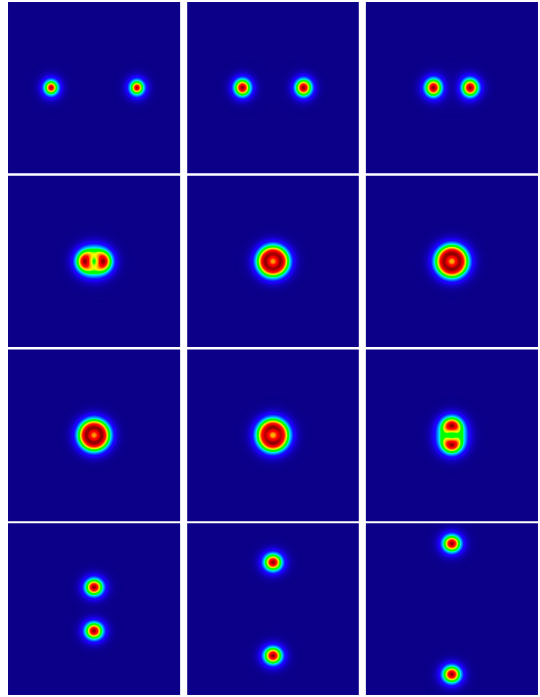}

\caption{
Time slices of the energy density for the head-on ($b=0$) 
collision of two $m=1$ vortices with supercritical velocities for $\kappa=1$. 
The plots represent times $t=0, 5, 10, 16.75, 25, 50, 75, 91.75, 100, 107.5, 117.5,\ \rm{and} \ 125$,
displayed from top left to bottom right, respectively.
The simulations were conducted on a $401\times 401$ grid with domain spanning 
$\{x:-10<x<10\}$ and 
$\{y:-10<y<10\}$.
The initial velocities are tuned to within one part in $10^{13}$  of the critical solution,
and the long-lived $m=2$ static intermediate attractor can be observed ($25\lesssim t \lesssim 91.75$).
The solution demonstrates right-angle scattering, $\theta_s = 90^{\circ}$.  
\label{fig:rtangle}}
\end{figure}
%
%

%
The threshold of backscattering and right-angle scattering is first explored for $\kappa=1$.
Initial values of $v_0$ are selected that are observed to be above and below $v_0^*$ and are referred to as 
$v_0^+$ and $v_0^-$, respectively.  
Vortices colliding with velocity $v_0^+$ exhibit right-angle scattering, while vortices with velocity $v_0^-$ backscatter.
A bisecting procedure  was performed to bracket $v_0^*$ with $v_0^\pm$ values that become successively closer
to $v_0^*$ until machine (double) precision is reached.
As $v_0$ approaches $v_0^*$ from either side, one observes longer interaction; the two $m=1$ vortices 
become a single $m=2$ vortex with a lifetime that grows as $v_0$ gets closer to $v_0^*$.
Figures \ref{fig:reflect} and \ref{fig:rtangle} demonstrate the time evolution of head-on vortex scattering simulations
for $v_0$ just below and above $v_0^*$, referred to as subcritical and supercritical scattering, 
respectively.\footnote{This ($v_0 = v_0^*$) critical scattering should not be confused with critical coupling 
($\kappa=1/\sqrt{2}$) of noninteracting vortices.}
%
%

\begin{figure}[t]
\includegraphics[width=9cm,height=16cm]{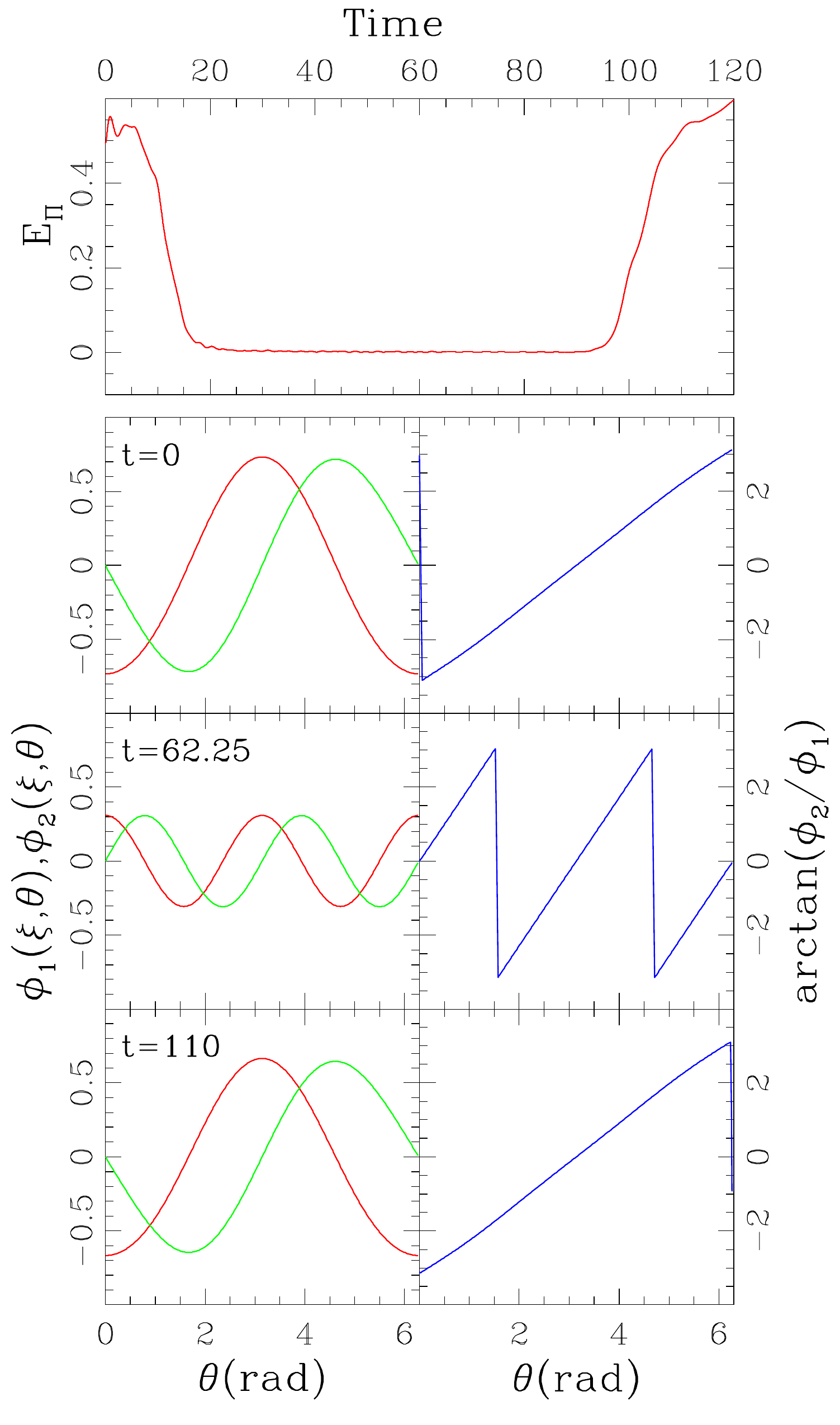}
\caption{
%
%
Plots demonstrating the static intermediate attractor for a near-critical solution.
The top plot demonstrates the dynamic energy of the scalar field ($E_\Pi$) as a function of time.
The lower plots show the real scalar field components $(\phi_1,\phi_2)$ (left) and the phase
(right) around the vortex at a radius equal to the coherence length, $\xi$, at three times:
before interacting ($t=0$), during the interaction ($t=62.25$), and after interacting ($t=100$).
It is clear that during the interaction, the scalar field is static ($E_\Pi\approx 0$) and %
has winding number $m=2$.
For before and after solutions, the left vortex is shown.
\label{fig:PhaseStatic}}
\end{figure}

Figure \ref{fig:reflect} demonstrates subcritical scattering with $v_0$ tuned to  one part
in $10^{13}$ below $v_0^*$.
The vortices are observed to collide, overcoming the mutual repulsion, at $t\approx 17$.
For times $17 \lesssim t \lesssim 92$, the $m=2$ vortex is approximately static, and at
$t\approx 92$, the $m=2$ vortex decays into two $m=1$ vortices with $\theta_s = 180$.
Similarly, Fig. \ref{fig:rtangle} demonstrates supercritical scattering with $v_0$ tuned to  one part
in $10^{13}$ above $v_0^*$.
%
The vortices are observed to collide, to overcome mutual repulsion, 
to coalesce into an $m=2$ vortex,
and to eventually decay into two $m=1$ vortices, 
only this time with  $\theta_s = 90$.
Based on conservation of topological charge, 
the presence of an approximate 
$m=2$ vortex was well understood to exist previously \cite{SHELLARD1988262}.
However, the ability to fine-tune the collision to allow a static 
$m=2$ vortex to be created and exist for an arbitrary amount of time 
is new to this work.

To confirm the presence of an $m=2$ static vortex as the intermediate attractor, 
it is useful to consider the dynamic energy of the scalar field,
\begin{equation}
E_\Pi = \frac{1}{2\kappa^2}\int dx dy \left(  \Pi_1^2 + \Pi_2^2 \right),
\end{equation}
which is defined here  simply to be contribution of the kinetic energy of the scalar field from only the time derivative terms, i.e., 
without the ``shape" energy from the spatial derivative terms.
Figure \ref{fig:PhaseStatic} demonstrates the dynamic energy in the scalar field as a function of time for the same
subcritical collision shown in Fig. \ref{fig:reflect}.
Figure \ref{fig:PhaseStatic} also shows the components of the complex scalar field (left) and the phase of the scalar field (right)
along a circular path centered on the vortex with radius equal to the coherence length, $\xi=1$, 
for  before, during, and after the collision.
At times $t=0$ and $t=110$, the phase of the scalar field rotates once ($m=1$) around the left vortex, 
whereas at $t=62.25$, the phase rotates twice ($m=2$) around the single  vortex at the origin.
While this result is not surprising, it definitively shows that the intermediate state during the deep inelastic scattering 
is indeed a vortex with winding number $m=2$.
%

\begin{figure}[t]
\includegraphics[width=0.95\linewidth]{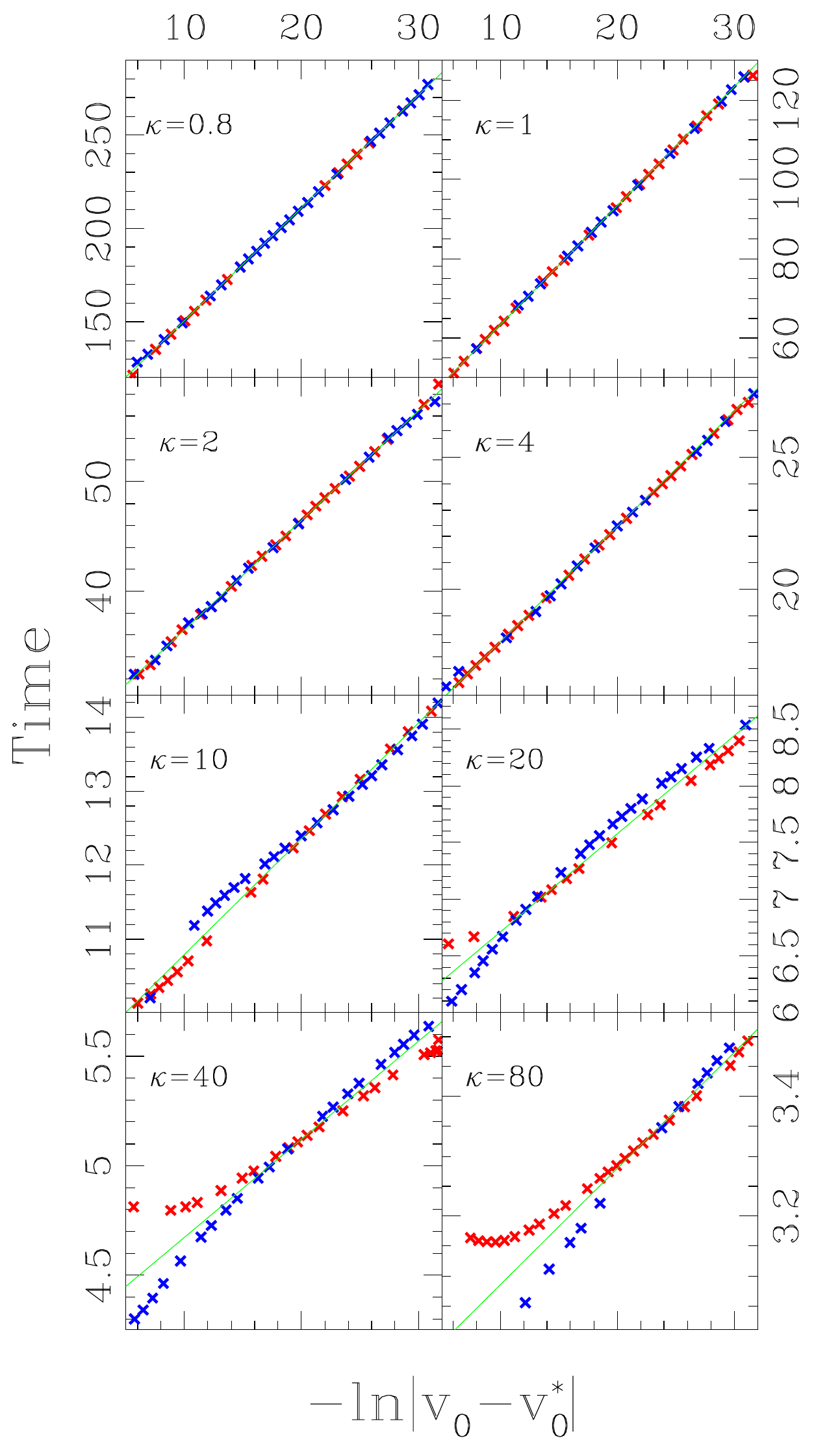}
\caption{
Solution lifetime as a function of $-\ln |v_0 - v_0^*|$
 for a range of $\kappa$.  
Time-scaling laws are observed as the critical solution is approached in parameter space.
The critical solutions and associated scaling exponents, $\alpha$ (slope of green best-fit line), 
are $\kappa$ dependent.
Solutions are colored red or blue for evolutions with $v_0$ above or below $v_0^*$, respectively.
%
%
\label{fig:TimeScaling}}
\end{figure}

Figures \ref{fig:reflect}, \ref{fig:rtangle}, and \ref{fig:PhaseStatic} also begin to show the longevity of the  
$m=2$ vortex and its role as an intermediate attractor.  
The upper-right graph of Fig. \ref{fig:TimeScaling} demonstrates the ``lifetime" of each collision as a function 
of $\ln|v_0-v_0^*|$ for $\kappa=1$.  
Since all of the evolutions for a given $\kappa$ start out with vortices in the same positions 
and end based on common exit criteria,
the  relative increase in lifetime is a good measure 
of the lifetime of the $m=2$ intermediate vortex.
It is clear that a time-scaling relationship is observed,
\begin{equation}
T \propto - \alpha \ln \left|v_0-v_0^*\right|,
\end{equation}
where the slope $\alpha$ governs the scaling and the $y$ intercept is ignored because it depends on the initial positions
and exit criteria.
This type of time scaling has been observed in many contexts, including 
the gravitational collapse of Yang-Mills and boson fields \cite{PhysRevLett.77.424,PhysRevD.56.R6057,PhysRevD.62.104024}
and the self-interaction of nonlinear oscillons in flat space \cite{PhysRevD.52.1920,PhysRevD.65.084037,PhysRevD.100.116005}.
In each of these cases, the time scaling describes the life of an unstable intermediate attractor solution that can be 
observed by tuning a parameter that results in  
one of two possible  outcomes.  
In the case of gravitational collapse, the threshold is the creation (or not) of a black hole.
In the case of oscillons, the threshold appears at resonances that separate solutions with $n$ and $n+1$ 
``shape mode" modulations.  
In this work, the discrete outcomes are right-angle and backward scattering resulting from the head-on ($b=0$) 
collision of two $m=1$ vortices.

While the above discussion is for $\kappa=1$ vortex collisions, there exist similar time-scaling relations for
all of the repulsive $(\kappa > 1/\sqrt{2})$ vortices explored.  
Figure \ref{fig:TimeScaling} demonstrates the scaling laws for a wide range of $\kappa$, from $\kappa=0.8$ to $\kappa = 80$.
This dynamic range in $\kappa$ 
corresponds to $ 2.56 \leq \lambda_{\rm MRS} \leq 25,600$  in the Abelian-Higgs model studied in \cite{PhysRevD.45.1355}, 
where the authors explored  $1 \leq \lambda_{\rm MRS} \leq 32$. 
%
\begin{table}[b]
\begin{tabular}{ S[table-format=4.4] c  S[table-format=3.5]  | c S[table-format=5.1]  S[table-format=4.2]  }
\hline
\hline
$\kappa$ &   $v_0^*$  &  $\alpha$   & & $Nx$  & $x_{\rm max}$  \\
\hline
0.8	&	0.152		&	6.03		& & 401 	&	10.0 	\\
1.0	&	0.271		&	3.00		& & 401	& 	10.0  \\
2.0	&	0.482		&	1.00		& & 587	& 	10.92  \\
4.0	&	0.617		&	0.436	& & 757	& 	8.06  \\
8.0	&	0.706		&	0.195	& & 1005	& 	5.73  \\
10.0	&	0.728 		&	0.150	& & 1733	& 	8.00  \\
20.0	&	0.790		&	0.0897	& & 1517	& 	3.58  \\
40.0	&	0.814		&	0.0448	& & 2088	& 	2.48 \\
80.0	&	0.837 		&	0.0202	& & 2773	& 	1.65 \\
\hline
\end{tabular}
\caption{
Table of observed critical velocities and time-scaling exponents for various $\kappa$.
Critical velocities were fine-tuned to machine (double) precision but are reported here to 
three significant figures.
Simulation parameters are provided, and all grids are square with $Nx=Ny$,
$\{x:-x_{\rm max}\leq x \leq x_{\rm max}\}$, and 
$\{y:-y_{\rm max}\leq y \leq y_{\rm max}\}$, where $y_{\rm max} = x_{\rm max}$.
\label{table:SimulationParameters}
}
\end{table}
%
Table \ref{table:SimulationParameters} captures the results 
for the critical velocities, time-scaling exponents,  grid size, and dimensions for
each $\kappa$ explored.

\begin{figure}
\includegraphics[width=0.9\linewidth]{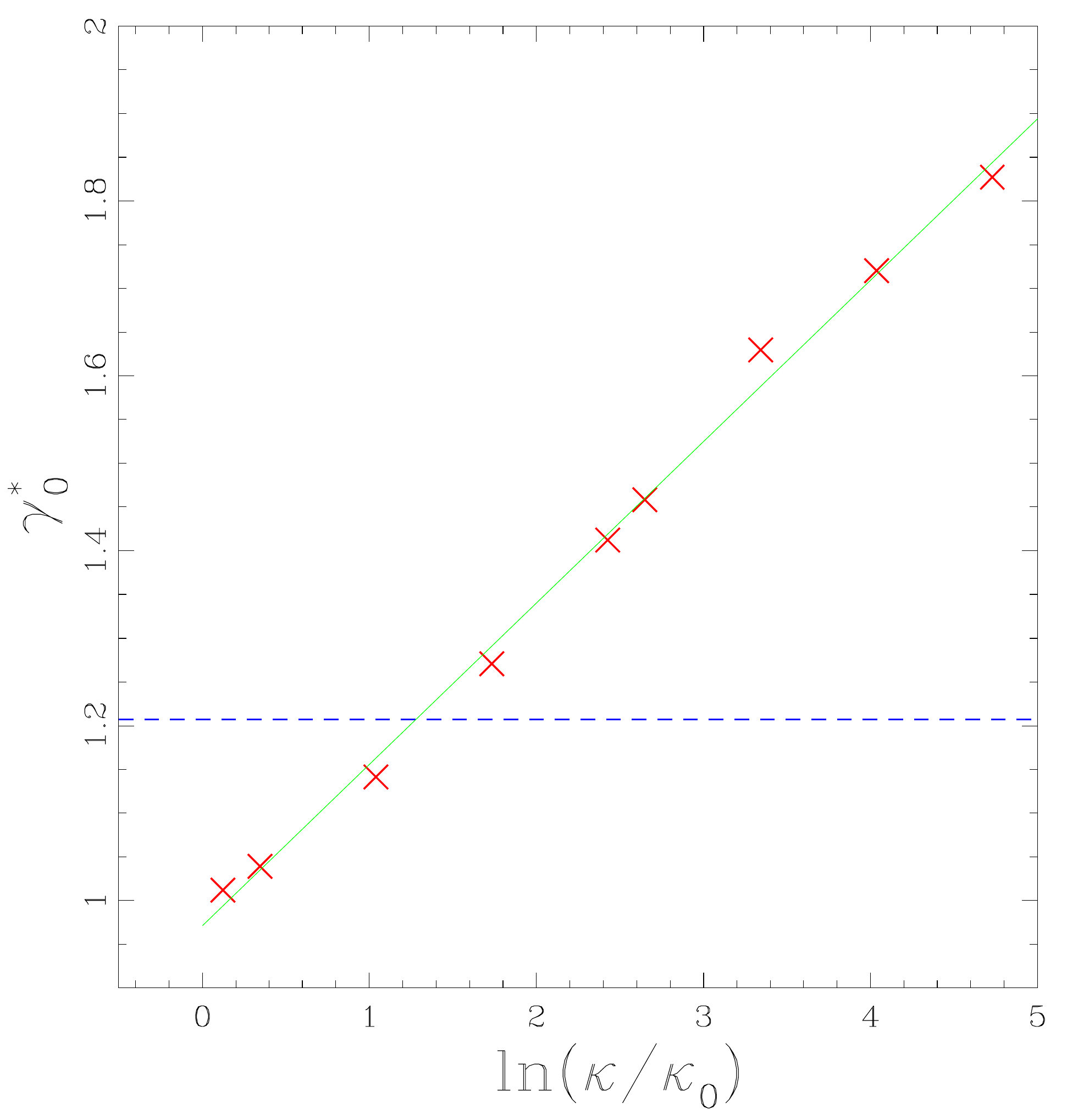}%
\caption{
Plot of $\gamma_0^* = \left( 1- (v_0^*)^2\right)^{-1/2}$ dependence on ${\rm ln}(\kappa)$.   
Red $\times$'s denote $\gamma_0^*$ for each critical solution obtained, 
and the green line is the best fit for a linear dependence, 
$\gamma_0^* \propto m_\gamma {\rm ln}(\kappa / \kappa_0)$, 
where $m_\gamma \approx 0.184$. 
%
The blue dashed line represents $\gamma_0^* = 1.21$, corresponding to the previously understood upper bound 
of $v_0^*(\kappa)\approx 0.56$.
\label{fig:gamma_vs_logkk0}}
\end{figure}

One observation that can immediately be made from Table \ref{table:SimulationParameters} is that the previously 
understood upper bound on critical velocity of $v_0^*\approx 0.56$ does not exist. 
 Figure \ref{fig:gamma_vs_logkk0} clearly demonstrates a relationship 
between ${ \gamma_0^* =  \left( 1- (v_0^*)^2\right)^{-1/2}}$ and the 
Ginzburg-Landau parameter, 
\begin{equation}
\gamma_0^* = m_\gamma \ln \left(\kappa/\kappa_0\right) + c_\gamma,
\end{equation} 
where the best fit gives $m_\gamma = 0.184$, $c_\gamma = 0.97$, and where $\kappa_0 \equiv 1/\sqrt{2}$.
While the factor of $\kappa_0$ is redundant and could be absorbed into $c_\gamma$, 
this form nicely results in the relevant domain of $\ln(\kappa/\kappa_0)$ being $[0,+\infty]$.
The observed value of $c_\gamma \approx 1$ also suggests that 
\begin{equation}
\left( \gamma_0^* -1\right) \approx m_\gamma \ln \left(\kappa/\kappa_0\right), 
\end{equation}
which would imply that as $\kappa$  approaches $\kappa_0$,  $v_0^*$ approaches zero. 
This clear relationship not only provides new insight into vortex scattering, 
but it also provides convincing evidence that the only upper bound to $v_0^*(\kappa)$ would be the speed of light,
and this would occur in the limit of infinite $\kappa$.
It is this author's belief that the work of \cite{PhysRevD.45.1355} was groundbreaking and incredibly thorough 
for its time, but it is likely that the simulations performed therein 
 were  underresolved.
The higher ratio of vortex size to lattice spacing used here  for the entire range of $\kappa$ being explored appears to be necessary 
to fully resolve vortex-vortex interactions.
%

\begin{figure}[t]
\includegraphics[width=0.9\linewidth]{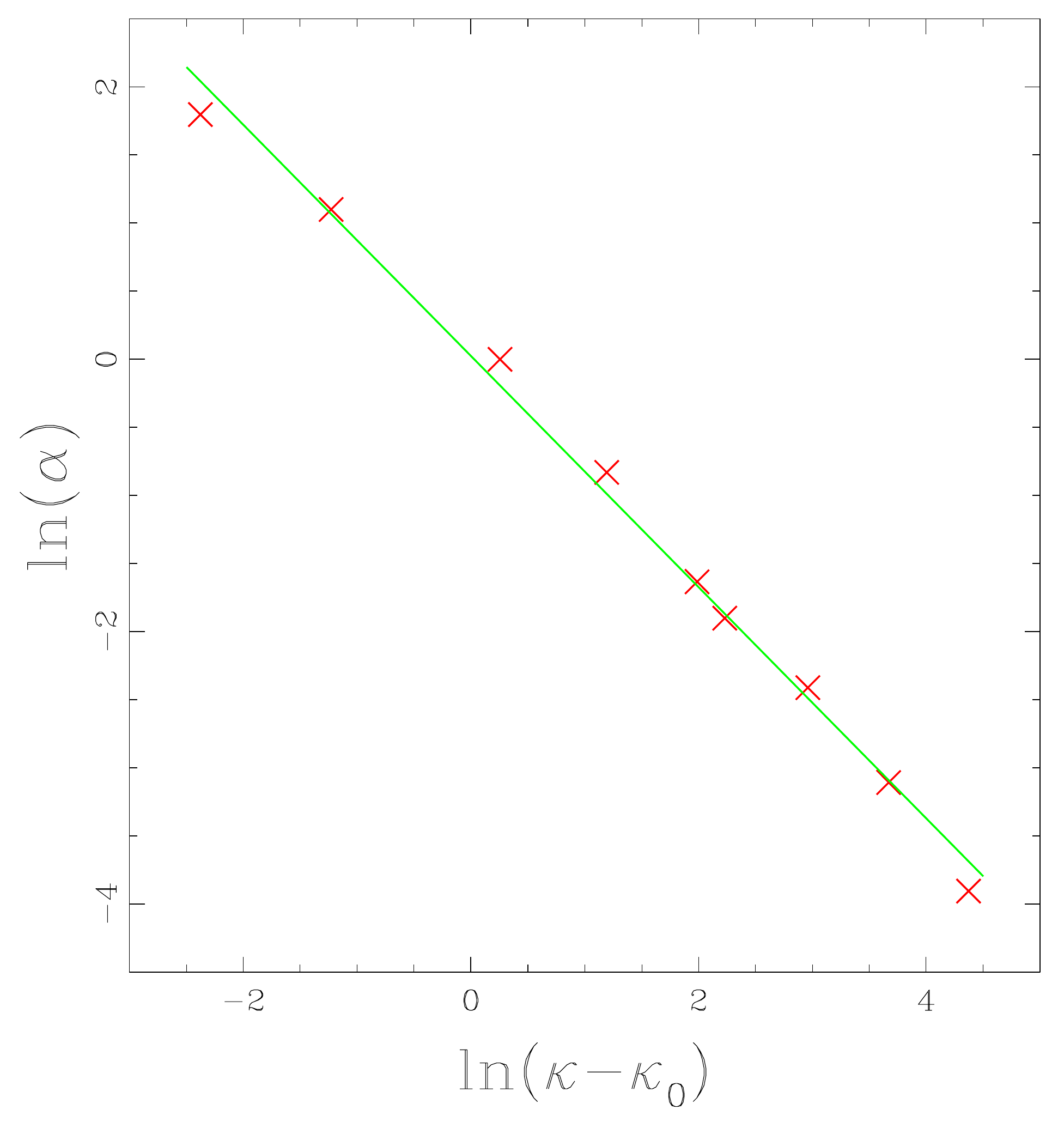}%
\caption{
Plot of ${\rm ln}(\alpha)$ dependence on $ {\rm ln}(\kappa-\kappa_0)$.  
Red $\times$'s denote $\ln(\alpha)$ for each critical solution obtained, 
and the green line is the best fit for a linear dependence, 
${\rm ln}(\alpha) \propto m_\alpha  {\rm ln}(\kappa-\kappa_0)$, 
where $m_\alpha = -0.85$ and $\kappa_0 = 1/\sqrt{2}$. 
%
%
%
\label{fig:KappaSlope}}
\end{figure}

Another observation one can make from Table \ref{table:SimulationParameters} (and Fig. \ref{fig:KappaSlope})
 is that there appears to be a relationship
between the time-scaling exponent and the Ginzburg-Landau parameter,
%
\begin{equation}
\ln \alpha =  m_\alpha \ln \left(\kappa -\kappa_0\right)+ c_\alpha,
\end{equation}
where the best fit gives $m_\alpha = -0.85$, $c_\alpha = 0.02$, and again, where $\kappa_0 \equiv 1/\sqrt{2}$.
This suggests that the 
 time-scaling exponent, $\alpha$, is (nearly) inversely proportional to the Ginzburg-Landau parameter, $\kappa$,
\begin{equation}
\alpha \approx \left( \kappa - \kappa_0\right)^{-0.85}.
\end{equation}
In the limit where $\kappa$ approaches $\kappa_0$, the vortices become less and less repulsive, the critical velocity
goes to zero, and the lifetime of the intermediate attractor goes to infinity.  This is actually just the trivial case of 
two noninteracting vortices located at the same position at rest.  
In the limit where $\kappa$ approaches infinity, the time-scaling exponent goes to zero.
This results in 
shorter 
lifetimes for larger $\kappa$, as seen in Fig. \ref{fig:TimeScaling}.
This  should also be intuitively expected in that with larger $\kappa$, the vortices are more repulsive, 
and it becomes harder to suppress (through fine-tuning $v_0$)  the increasingly dominant unstable mode, where 
the scaling exponent $\alpha$ is inversely proportional to the Lyapunov exponent of the unstable mode, 
similar to \cite{gundlach1997critical,hara1996renormalization}.

These results not only adjust the previous understanding of  head-on ($b=0$) vortex collisions  (the  
upper bound of $v_0^*$), 
but they also provide additional insight into head-on collision dynamics that were not previously understood.  
It is definitively shown that the $m=2$ vortex is the intermediate attractor on the threshold of right-angle scattering, and
more interestingly, the new scaling relations observed here strongly suggest that with appropriate fine-tuning of $v_0$, 
one can predictably create an arbitrarily long-lived $m=2$ vortex from the head-on collision of two $m=1$ vortices.
%

\section{Off-axis Scattering \label{sec:OffAxis}}

\begin{figure}[t]
\includegraphics[width=0.9\linewidth]{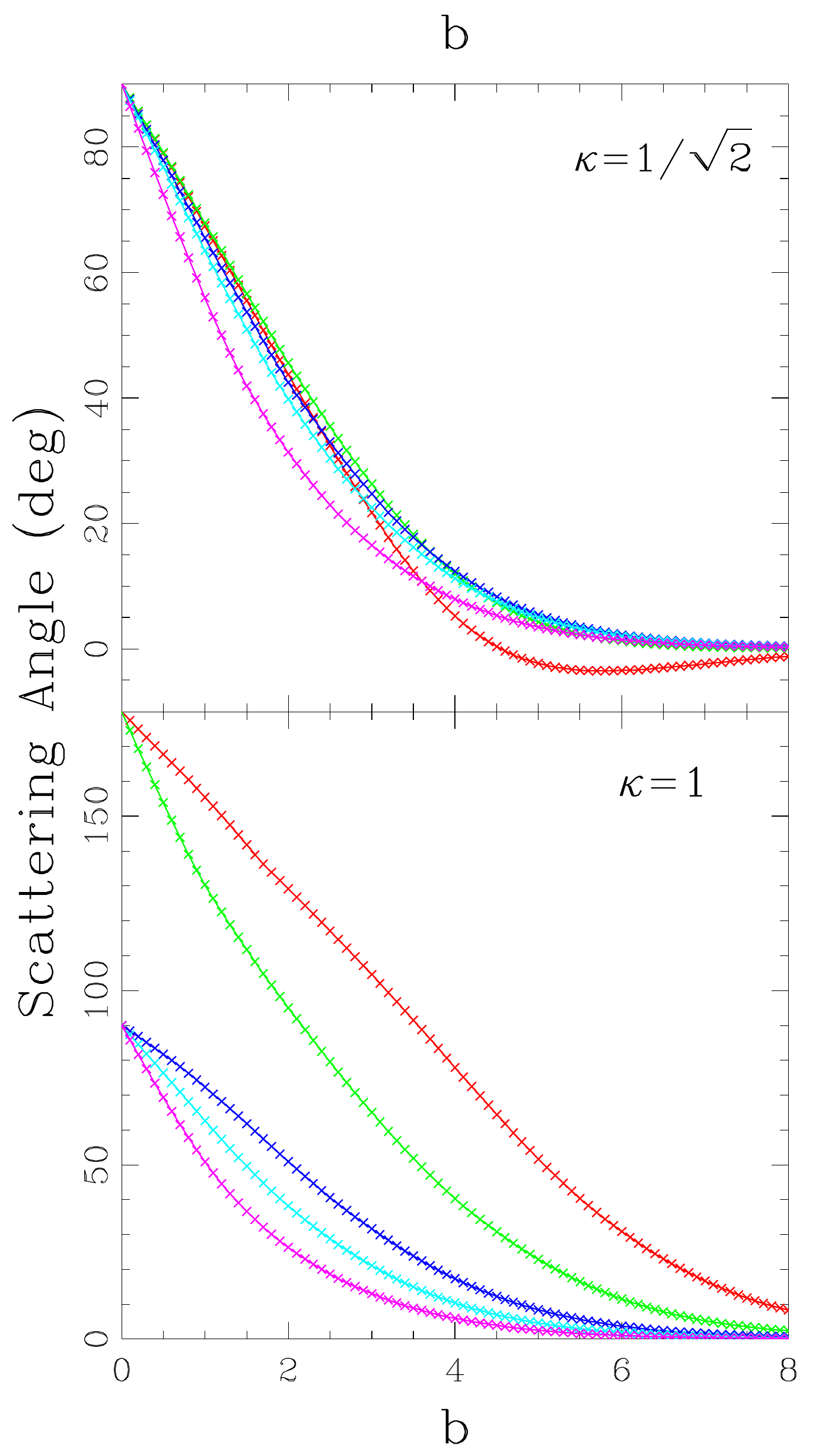}%
\caption{
Plots of scattering angle dependence on impact parameter, $b$, for $\kappa = 1/\sqrt{2}$ (top) 
and $\kappa=1$ (bottom) for $v_0=0.1, 0.2, 0.4, 0.6, \  {\rm and} \ 0.8$, plotted in red, green, blue, cyan, and magenta,
respectively.
noninteracting vortices ($\kappa = 1/\sqrt{2}$) demonstrate 
right-angle scattering only, while $\kappa=1$ vortices demonstrate both right-angle scattering and backscattering, 
depending on the collision velocity ($v_0^* \approx 0.271$). 
\label{fig:DeflectionLinear}}
\end{figure}

%
This section presents new results for off-axis 
scattering of $m=1$ vortices.
Initial data are again prepared as described in Sec. \ref{sec:InitialDataEOM}, 
only now for $b\neq 0$.
%
%
Figure \ref{fig:DeflectionLinear} shows results of scattering angle as a function of impact parameter, $\theta_s(b)$,
for $\kappa=1/\sqrt{2}$ (top) and $\kappa=1$ (bottom) with linearly sampled $b$, 
and where each $\theta_s(b)$ curve is for a given $v_0$.
The $\theta_s(b)$  for noninteracting vortices 
converge to a single point, $\theta_s = 90^\circ$,  at $b=0$,
whereas the $\theta_s(b)$  for repulsive vortices converge to either 
$\theta_s = 90^\circ$ or $\theta_s = 180^\circ$ at $b=0$, 
depending on whether $v_0$ is above or below the critical  velocity ($v_0^* \approx 0.271$ for $\kappa =1$), respectively. 
Figure \ref{fig:DeflectionLinear} displays  $\theta_s(b)$ with 81 points per $\theta_s(b)$ for 5 different $v_0$.  
As such,
these data 
enhance the results of \cite{SHELLARD1988262} for noninteracting ($\lambda_{\rm SR}=1$) collisions 
and  
\cite{PhysRevD.45.1355} for repulsive ($\lambda_{\rm MRS}=4$) vortices
and 
serve as a starting point to better understand the off-axis
deep inelastic scattering of vortices.
%
\begin{figure}[t]
\includegraphics[width=0.9\linewidth]{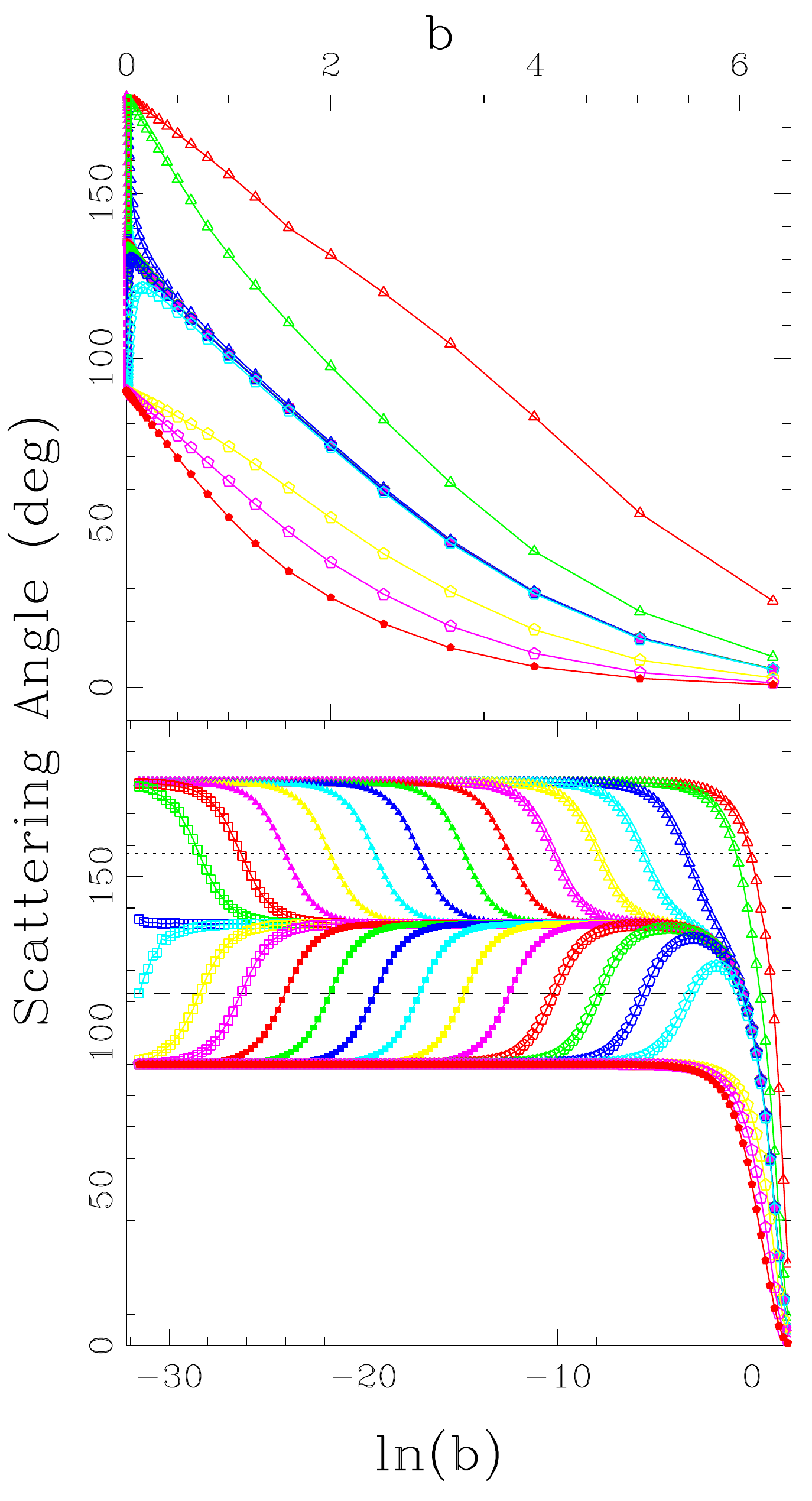}%
\caption{
Plots of scattering angle dependence on impact parameter, $b$,
for $\kappa=1$, shown for linear $b$ (top) and $\ln(b)$
(bottom).
Plots display logarithmically sampled $b$
for $v_0 = 0.1, 0.2, 0.4, 0.6, 0.8, \  {\rm and} \  (1\pm 10^{-n})\cdot v_0^*$, for $n=2$ through $n=14$ at 
an increment of $\Delta n = 1$.
Plots are shown in increasing $v_0$ with open then closed triangles, squares, and pentagons, each cycling through the colors
red, green, blue, cyan, yellow, and magenta, respectively.
The dotted and dashed lines represent $\theta_s=157.5^\circ$ and $\theta_s = 112.5^\circ$, respectively.  
Intersections of the plots with the dotted and dashed $\theta_s$-constant lines 
denote departures from $\theta_s=135^\circ$ scattering. 
%
\label{fig:DeflectionLog}}
\end{figure}

%
\begin{figure}
\includegraphics[width=0.95\linewidth]{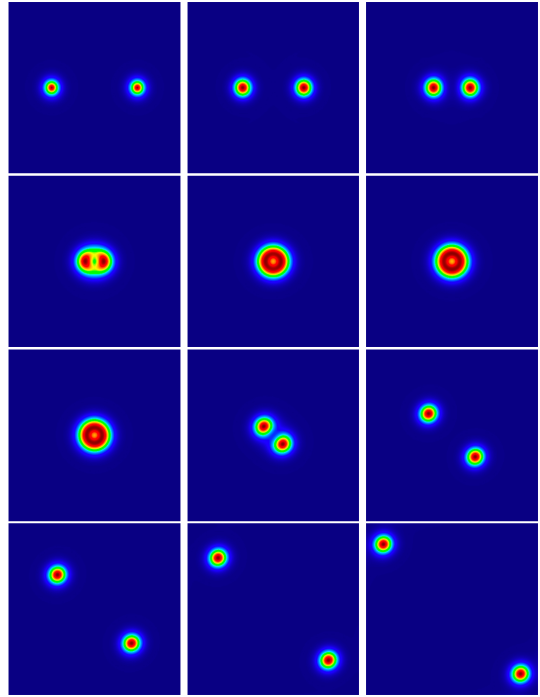}

\caption{
Time slices of the energy density for the nearly head-on ($b=10^{-10}$) 
collision of two $m=1$ vortices with supercritical velocities for $\kappa=1$. 
The plots represent times $t=0, 5, 10, 16.75, 25, 50, 75, 91.75, 100, 107.5, 117.5,\ \rm{and} \ 125$,
displayed from top left to bottom right, respectively.
The simulations were conducted on a $401\times 401$ grid with domain spanning 
$\{x:-10<x<10\}$ and 
$\{y:-10<y<10\}$.
The initial velocities are tuned to within one part in $10^{13}$  of the critical solution,
and the long-lived $m=2$ static intermediate attractor can be observed ($25\lesssim t \lesssim 75$).
The solution demonstrates scattering with $\theta_s = 135^{\circ}$.  
\label{fig:offaxis_bp}}
\end{figure}

By observing  $\theta_s(b)$ curves with $v_0$ approaching $v_0^*$, 
one can explore the threshold of right-angle scattering  by using two 
 dynamical degrees of freedom, $b$ and $v_0$.
Figure \ref{fig:DeflectionLog}  displays $\theta_s(b)$  for $\kappa = 1$, with  $b$ sampled logarithmically to
more easily observe the  behavior in the $b\rightarrow 0$ limit.
To  allow Fig. \ref{fig:DeflectionLinear} (bottom) and previous work  \cite{PhysRevD.45.1355} to be  easily compared
 to Fig.  \ref{fig:DeflectionLog}, the top graph in Fig.  \ref{fig:DeflectionLog} displays  data as a function of linear $b$ 
while the bottom graph displays the same data as a function of $\ln(b)$.
While the same $v_0$ used in Fig. \ref{fig:DeflectionLinear} are used in Fig. \ref{fig:DeflectionLog}, 
an additional 26 values of $v_0$ are also used that
logarithmically approach $v_0^*$ from above and below in constant intervals of $\ln|v_0-v_0^*|$,
\begin{equation}
v_{0,n} = \left( 1 \pm 10^{-n}\right)  v_0^*,
\end{equation}
where $n$ ranges from 2 to 14 with $\Delta n = 1$,
and the $b$ are sampled at 10 samples per decade for a total of 141 points for each $\theta_s(b)$,
resulting in data from 4371  vortex-vortex collisions.
It is immediately apparent that there exists more structure in the $\theta_s(b)$ curves than was originally observed 
in
Fig. \ref{fig:DeflectionLinear} and \cite{SHELLARD1988262,PhysRevD.45.1355}.
%
%
Figure  \ref{fig:DeflectionLog} clearly demonstrates 
that as $v_0$ approaches $v_0^*$ from above or below, 
the $\theta_s(b)$ ``latch on" to an intermediate attractor solution
that plateaus at $\theta_s \approx 135^\circ$.
Figure \ref{fig:offaxis_bp} shows the time evolution of a slightly off-axis collision ($b=10^{-10}$) for $v_0$ tuned to
$v_0^*$ to about one part in $10^{13}$.
Although $\kappa=1$ solutions will always have either  $\theta_s=90^\circ$ or $\theta_s=180^\circ$
at exactly $b=0$,
as $v_0 \rightarrow v_0^*$
the $b$ at which $\theta_s(b)$ departs from $\theta_s \approx 135^\circ$ (with decreasing $b$) 
approaches zero.
%
%

\begin{figure}[t]
\includegraphics[width=0.9\linewidth]{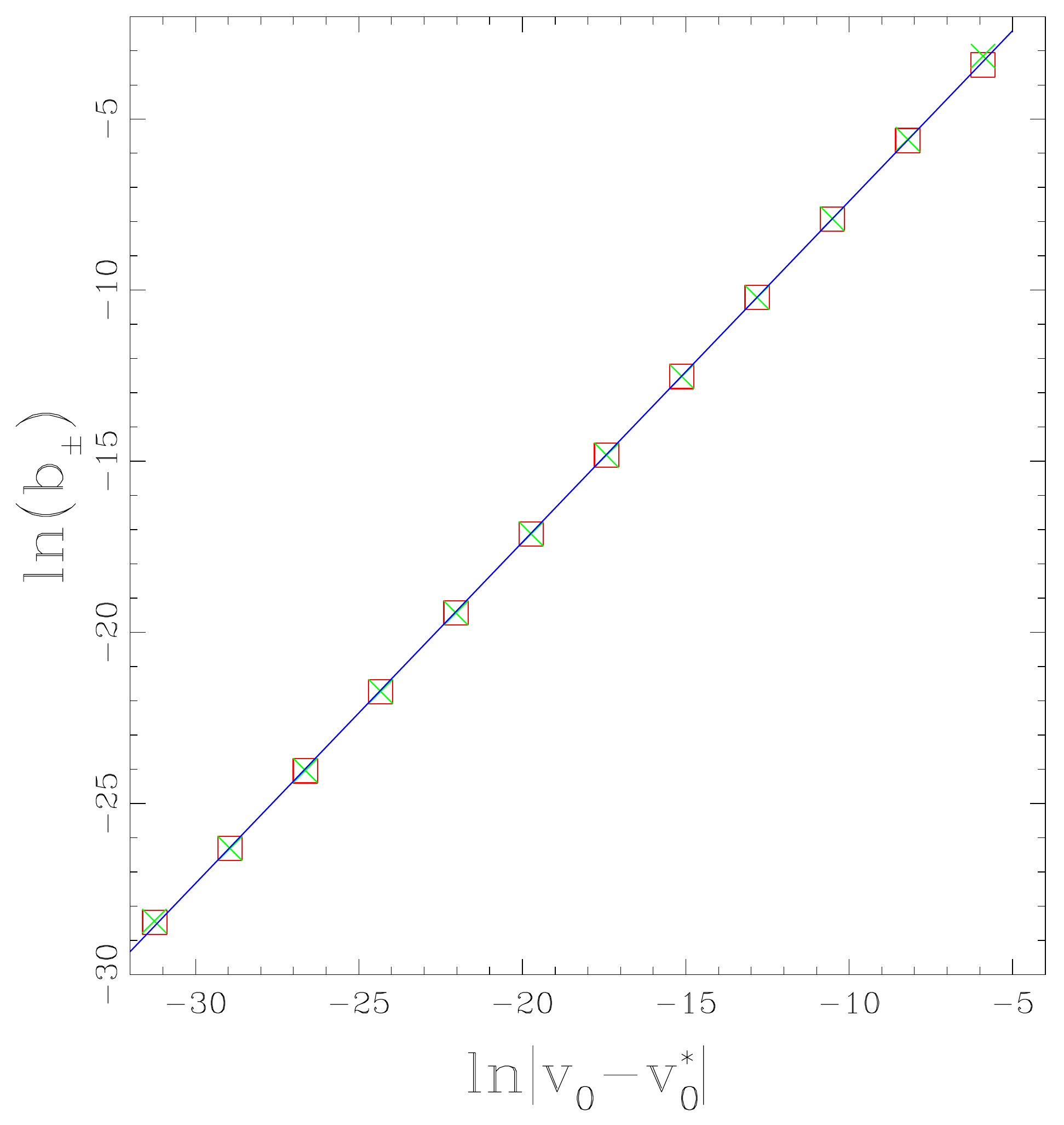}%
\caption{
Plot of $\ln(b_\pm)$  dependence on $\ln|v_0 - v_0^*|$ for $\kappa = 1$ and a range of $v_0$ approaching
the critical velocity, $v_0^*$, from above and below. 
As $v_0$ approaches $v_0^*$, the impact parameter denoting a transition from 
$\theta_s = 135^\circ$ scattering to $\theta_s = \{90^\circ,180^\circ\}$ scattering ($b_\pm$)  goes to zero.
The $b_+$ values are plotted with red squares and the $b_-$ values with green $\times$'s. 
The best-fit line is in blue and has slope $m_b \approx 1.00$ and $y$ intercept $c_b \approx 2.56$.
\label{fig:CriticalScatteringAngleKap1}}
\end{figure}
%

This  effect can be quantitatively analyzed by looking at the values of $b$ for which the $\theta_s(b)$ curves 
diverge from $\theta \approx 135^\circ$ as $b$ approaches zero. 
This divergence can be measured by finding the impact parameter at which $\theta_s(b)$ intersects the angle that is halfway between 
$\theta_s=135^\circ$ and its $\theta_s(b=0)$ value.  
For instance, subcritical collisions ($v_0 < v_0^*$) with $\theta_s(b=0) = 180^\circ$ are determined to diverge when
$\theta_s(b) = 157.5^\circ$, denoted by intersections with the dotted line in Figure \ref{fig:DeflectionLog} (bottom).   
Likewise,  supercritical collisions ($v_0 > v_0^*$) with $\theta_s(b=0) = 90^\circ$ are determined to diverge when
$\theta_s(b) = 112.5^\circ$, denoted by intersections with the dashed line in Figure \ref{fig:DeflectionLog} (bottom).   
These impact parameter values are denoted $b_\pm$, where 
$b_+$ is for subcritical solutions with $\theta_s(b_+) > 135^\circ$
and 
$b_-$ is for supercritical solutions with $\theta_s(b_-) < 135^\circ$.
Figure \ref{fig:CriticalScatteringAngleKap1} plots  $\ln (b_\pm)$ as a function of $\ln|v_0-v_0^*|$,
which clearly shows that as $v_0$ approaches $v_0^*$, the $b_\pm$ go to zero.  
Fitting the data to 
\begin{equation}
\ln(b_\pm) = m_b \ln|v_0 - v_0^*| + c_b \label{eqn:bpmscaling}
\end{equation}
results in a good fit to a straight line with best-fit  values of $m_b = 1.00$ and $c_b = 2.56$.
%

\begin{figure}[t]
\includegraphics[width=0.9\linewidth]{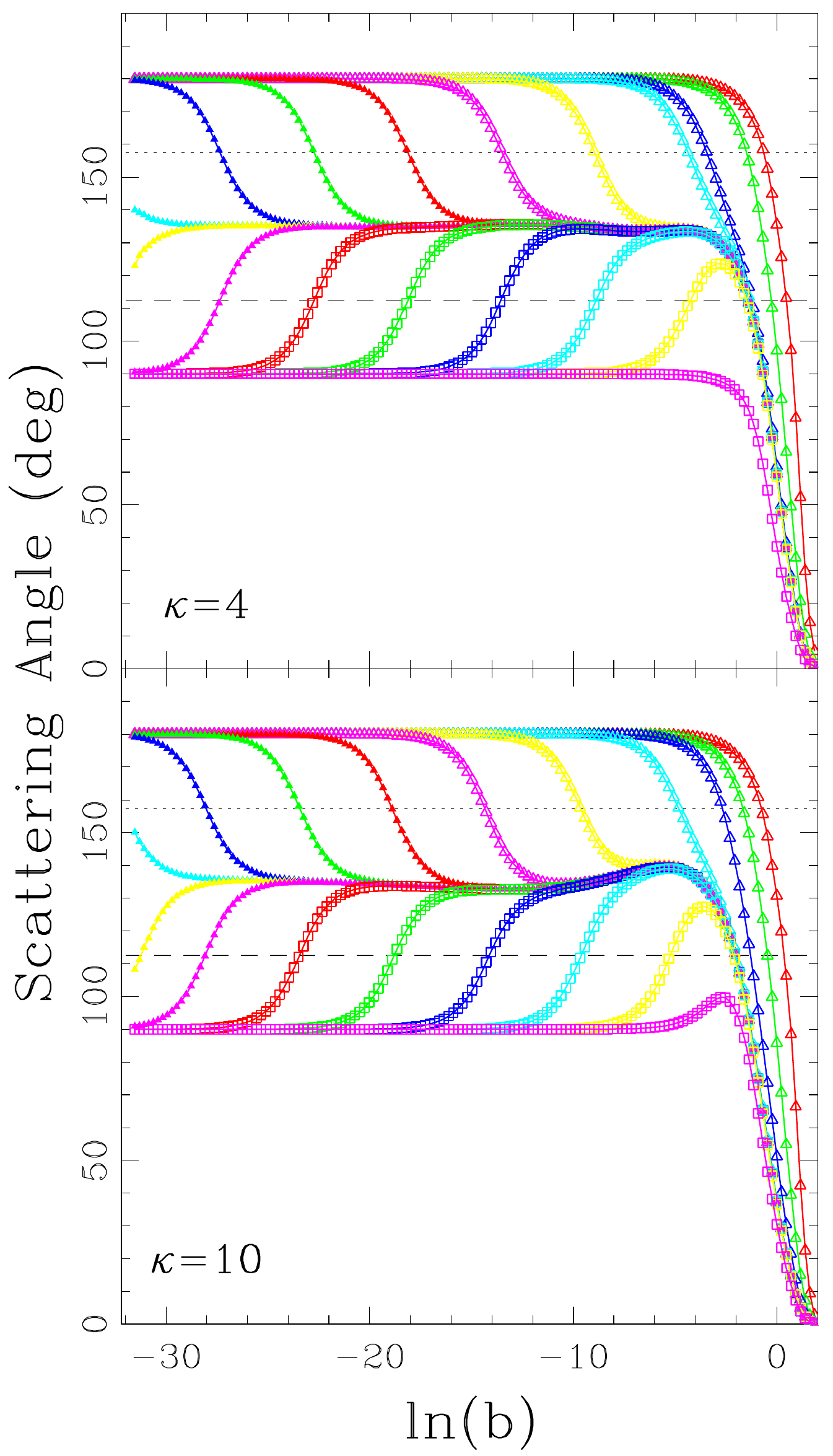}%
\caption{  
Plots of scattering angle  dependence on the logarithm of the impact parameter, $\ln(b)$,
for $\kappa = 4$ (top) and $\kappa = 10$ (bottom).
Plots display logarithmically sampled $b$
for $v_0 = 0.2, 0.4, 0.6, 0.8,\  {\rm and} \  (1\pm 10^{-n})\cdot v_0^*$, for $n=2$ through $n=14$ at 
an increment of $\Delta n = 2$.
Plots are shown in increasing $v_0$ with open then closed triangles and squares, each cycling through the colors
red, green, blue, cyan, yellow, and magenta, respectively.
The dotted and dashed lines represent $\theta_s=157.5^\circ$ and $\theta_s = 112.5^\circ$, respectively.  
Intersections of the plots with the dotted and dashed $\theta_s$-constant lines 
denote departures from $\theta_s=135^\circ$ scattering. 
\label{fig:CS_4_10_A}}
\end{figure}
\begin{figure}[t]
\includegraphics[width=0.9\linewidth]{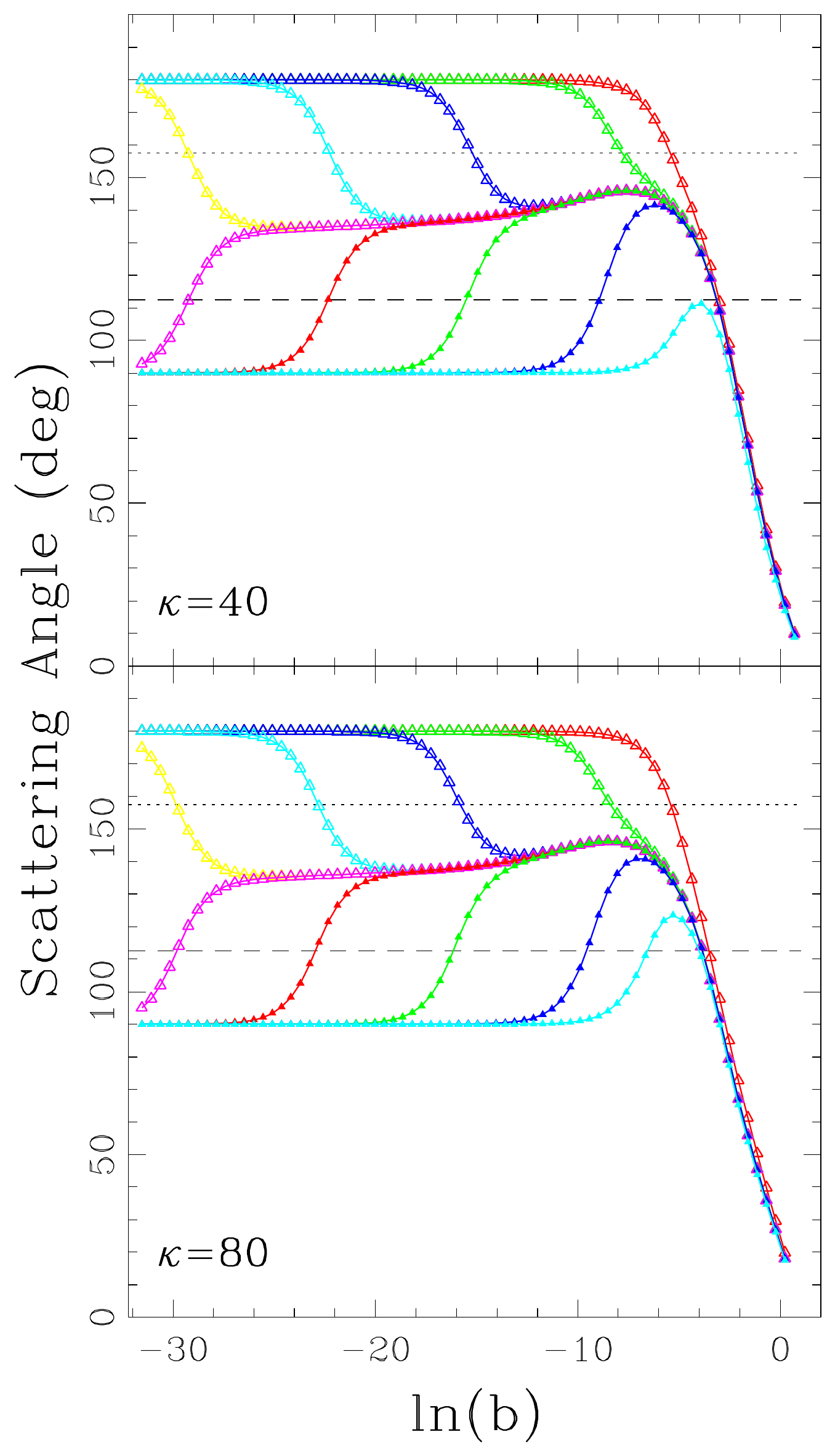}%
\caption{ 
Plots of scattering angle  dependence on the logarithm of the impact parameter, $\ln(b)$,
for $\kappa = 40$ (top) and $\kappa = 80$ (bottom).
Plots display logarithmically sampled $b$
for $v_0 = 0.8, 0.85,\  {\rm and} \  (1\pm 10^{-n})\cdot v_0^*$, for $n=3$ through $n=12$ at 
an increment of $\Delta n = 3$.
Plots are shown in increasing $v_0$ with open then closed triangles and squares, each cycling through the colors
red, green, blue, cyan, yellow, and magenta, respectively.
The dotted and dashed lines represent $\theta_s=157.5^\circ$ and $\theta_s = 112.5^\circ$, respectively.  
Intersections of the plots with the dotted and dashed $\theta_s$-constant lines 
denote departures from $\theta_s=135^\circ$ scattering. 
 \label{fig:CS_40_80}}
\end{figure}
\begin{figure}[t]
\includegraphics[width=\linewidth,height=\linewidth]{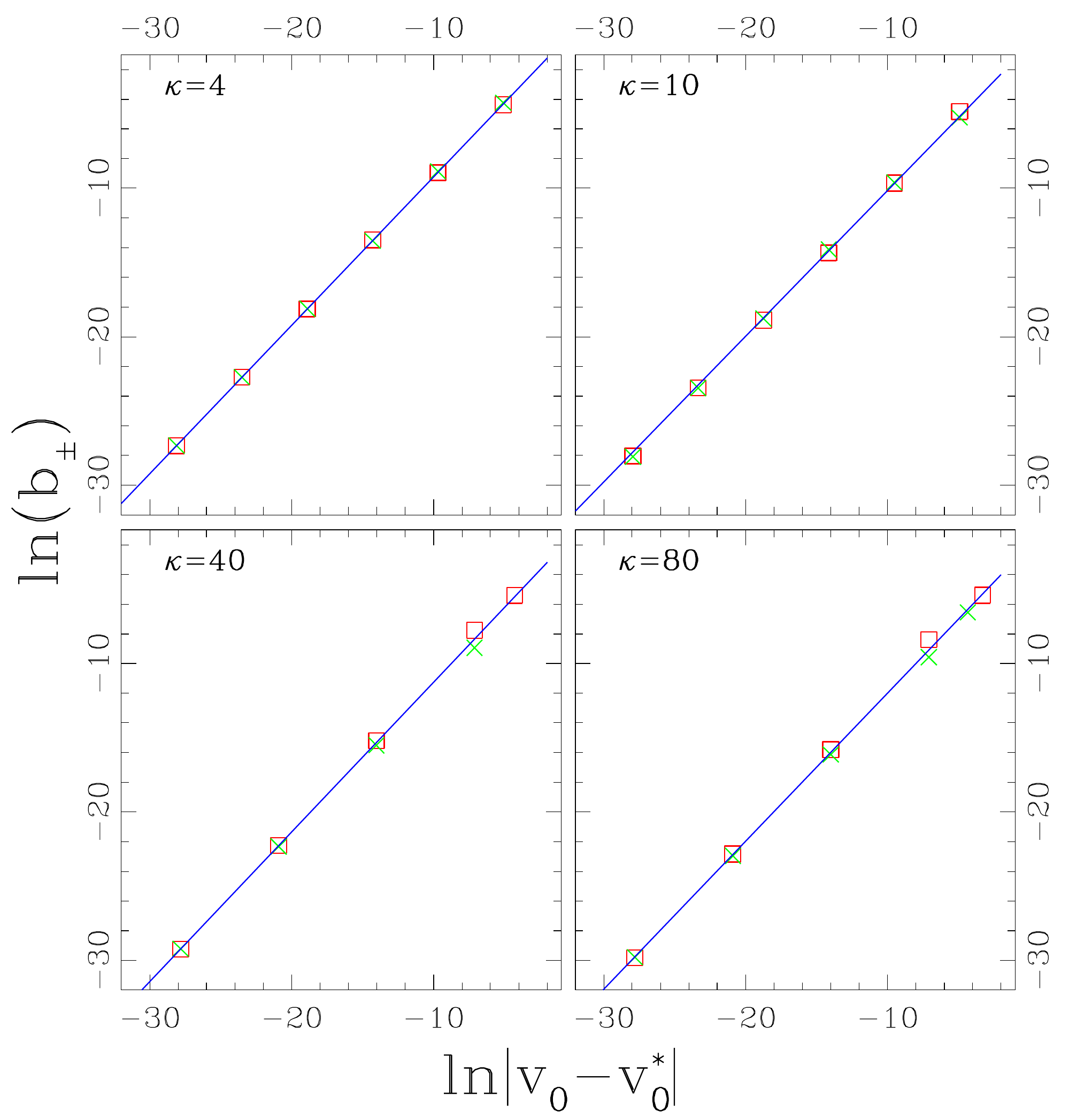}%
\caption{
Plot of $\ln(b_\pm)$  dependence on $\ln|v_0 - v_0^*|$ for $\kappa = 4, 10, 40, {\rm and} \ 80$
for a range of $v_0$ approaching the $v_0^*$ from above and below. 
As $v_0$ approaches $v_0^*$, the impact parameter ($b_\pm$) denoting a transition from 
$\theta_s = 135^\circ$  to $\theta_s = \{90^\circ,180^\circ\}$    goes to zero.
\label{fig:CriticalScatteringAngleKapAll}}
\end{figure}

Figures \ref{fig:CS_4_10_A} and \ref{fig:CS_40_80} show similar results for off-axis vortex collisions
with $\kappa=4, 10, 40, \ {\rm and}\  80$.
Figure \ref{fig:CS_4_10_A} shows $\theta_s(b)$ curves for $\kappa=4$ (top) and $\kappa=10$ (bottom),  
for $v_0 = 0.2, 0.4, 0.6, 0.8,\  {\rm and} \  (1\pm 10^{-n})\cdot v_0^*$, for $n=2$ through $n=14$ at 
an increment of $\Delta n = 2$,
and with $b$ sampled at 10 samples per decade, for a total of 2538 collisions for each $\kappa$. 
Figure \ref{fig:CS_40_80} shows $\theta_s(b)$ curves for $\kappa=40$ (top) and $\kappa=80$ (bottom),  
for $v_0 = 0.8, 0.85,\  {\rm and} \  (1\pm 10^{-n})\cdot v_0^*$, for $n=3$ through $n=12$ at 
an increment of $\Delta n = 3$,
and with $b$ sampled at five samples per decade, for a total of 710 collisions for each $\kappa$. 
While the density of $b$ samples and the number of $v_0$ explored are decreased with increasing $\kappa$
due to significantly increasing computational demands (while maintaining $\xi/dx > 10$), 
one begins to see common phenomena across
the entire range of $\kappa$ explored.

For all $\kappa$ observed, as $v_0$ approaches $v_0^*$, the $\theta_s(b)$ curves approach an intermediate attractor
solution that plateaus at $\theta_s = 135^\circ$ with $b_\pm$ that approach zero.
Figure \ref{fig:CriticalScatteringAngleKapAll} displays plots of $\ln(b_\pm)$ as a function of $\ln|v_0 - v_0^*|$
for $\kappa=4, 10, 40, {\rm and} \ 80$. 
It is again clear that there is a linear relationship and  (\ref{eqn:bpmscaling}) appears to accurately describe 
$b_\pm$, especially as $v_0$ approaches $v_0^*$.
Table \ref{table:CurveFits} captures the best-fit values for the range of $\kappa$ explored.
\begin{table}[b]
\begin{tabular}{S[table-format=5.3]  c S[table-format=5.4] }
\hline
\hline
$\kappa$ & $m_{b}$ & $c_b$  \\
\hline
1	& 	1.00		& 2.56 \\
4	&	1.00		& 0.79 \\
10	& 	0.98		& -0.32 \\
40 	& 	1.01		& -1.16 \\
80	& 	1.00		& -2.04 \\
\hline
\end{tabular}
\caption{
Best-fit parameters for relationship (\ref{eqn:bpmscaling}) for a range of $\kappa$.
Values are rounded to those in table. %
\label{table:CurveFits}
}
\end{table}
%
%
%
The numerical closeness of $m_b$ to 1 suggests the possibility that 
$m_b = 1$ exactly, which would give a simple  relationship,
\begin{equation}
b_\pm \approx c_b' |v_0 - v_0^*|, \label{eqn:bpmscalingapprox}
\end{equation}
where $c_b' = e^{c_b}$.

Independent of whether $m_b=1$ exactly, it has been shown that at the threshold of right-angle scattering,
$\theta_s(b)$  plateaus to $\theta_s=135^\circ$  and  $b_\pm\rightarrow 0$.
Since the $b_\pm$ at which $\theta_s(b)$ diverges  from $135^\circ$  goes to zero, 
$\theta_s(b)$ can be shown to be $135^\circ$ for an arbitrarily small $b$.
This behavior is demonstrated for a range of $\kappa$ and persists for arbitrarily small $b$, 
as long as $v_0$ is sufficiently fine-tuned to $v_0^*$.
So, while precisely head-on ($b=0$) scattering only allows $\theta_s=90^\circ$ or $\theta_s=180^\circ$, at criticality 
($v_0 = v_0^*$) there exists an 
intermediate attractor $m=2$ vortex solution that decays with a scattering angle of $\theta_s=135^\circ$
for arbitrarily small impact parameters.

\section{Conclusions}

Results have been presented from numerical simulations of the flat-space nonlinear Maxwell-Klein-Gordon 
equations with a spontaneously broken symmetry demonstrating deep inelastic scattering of $m=1$ 
vortices for a wide range of Ginzburg-Landau parameters ($\kappa$), impact parameters ($b$), 
and initial velocities ($v_0$).

Head-on ($b=0$) collision simulations were conducted that explored the threshold of right-angle scattering by varying 
the initial velocity. 
While the fact that repulsive vortices exhibit right-angle scattering above a critical velocity ($v_0^*$) has been known
for decades, until this work, the threshold of the transition from backward scattering to right-angle scattering
has not been thoroughly explored or understood.
This work corrects previous work that appeared to demonstrate an upper bound to the critical velocity.
In addition to this clarification, a new phenomenon was observed for $b=0$ collisions.
It was shown that solutions obey time-scaling laws, $T\propto \alpha \ln|v_0 - v_0^*|$, with $\kappa$-dependent 
scaling exponents $\alpha$ that vary according to a simple power-law dependence on $\kappa-\kappa_0$, 
where $\kappa_0 = 1/\sqrt{2}$ is the Ginzburg-Landau parameter for noninteracting vortices.
While expected, it was definitively shown that the intermediate attractor solution on the threshold is the $m=2$ vortex solution for
the $\kappa$ being investigated.

Off-axis ($b\neq 0$) collision simulations were conducted to gain new insight into vortex-vortex scattering, especially for small
impact parameters and with $v_0 \approx v_0^*$.
It is shown that for arbitrarily small (nonzero) impact parameters, the unstable intermediate attractor 
$m=2$ vortex decays with a $\kappa$-independent scattering angle of $135^\circ$.
This was demonstrated through careful analysis of scattering angle curves $\theta_s(b)$ and their divergence away from
$\theta_s = 135^\circ$ in the $v_0\rightarrow v_0^*$ and $b\rightarrow 0$ limits.  
%
Simply put, for an appropriately fine-tuned initial velocity, an arbitrarily small impact parameter can be shown 
to scatter at $\theta_s = 135^\circ$.
As such, this work has demonstrated the existence of a third outcome for ``head-on" ($b\rightarrow 0$) vortex collisions.

\appendix

\section{Dimensionless Lagrangian \label{app:DimensionLessLags}}

The Ginzburg-Landau system is described by the following 
dimensionful Lagrangian containing a  Maxwell field coupled to a complex scalar field in a gauge covariant way
with a spontaneously broken $U(1)$ symmetry:
\begin{eqnarray}
L
&=&  
-\frac{\epsilon_0 c^2}{4}F^{\mu\nu}F_{\mu\nu}  
-\frac{\hbar^2}{2m}g^{\mu\nu}D_\mu\phi 
                                                      \left( D_\nu\phi\right)^* 
+  \alpha_B  \phi^2 - \frac{\beta_B}{2}  \phi^4,   \nonumber \\
                                                \label{eqn:DimensionfulBECLagrangian}
\end{eqnarray}
where
\begin{eqnarray}
D_\mu\phi &=& \left( \partial_\mu  - i \frac{q}{\hbar} A_\mu\right)\phi,\\
F_{\mu\nu} &=& \partial_\mu A_\nu - \partial_\nu A_\mu,
\end{eqnarray}
 $A_\mu$ is the  Maxwell gauge potential,  $\phi$ is the order parameter, and $|\phi^2|$ 
is the number of superconducting particles per unit volume.
Following standard practice, to convert (\ref{eqn:DimensionfulBECLagrangian}) to a dimensionless Lagrangian,
one converts each coordinate and field variable to a dimensionless form,
%
\begin{eqnarray}
 \hat{x}^{\mu} 	&=& \left( \frac{1}{ \Lambda} \right) x^\mu, \\    
 \hat{A}_\mu 	&=& \left( \frac{ q \xi }{\hbar } \right) A_\mu, {\rm \ \ \ and} \\
 \hat{\phi}		&=& \left( \frac{\beta_B}{\alpha_B} \right)^{1/2} \phi, 
\end{eqnarray}
%
where one can use the standard definitions,
%
\begin{eqnarray}
\kappa				&=& \frac{\Lambda}{\xi}, \\
\frac{\alpha_B}{\beta_B} 	&=&  \phi_0^2, \\
\Lambda^{-2} 	&=& \frac{q^2 \phi_0^2 }{m \epsilon_0 c^2}, {\rm \ \ \ and} \\
\xi^2			&=& \frac{\hbar^2}{2 m \alpha_B} 
\end{eqnarray}
to obtain the dimensionless Lagrangian,
\begin{eqnarray}
\hat{L}
&=&  
-\frac{1}{4}\hat{F}^{\mu\nu}\hat{F}_{\mu\nu}  
-\frac{1}{2}g^{\mu\nu}\hat{D}_\mu\hat{\phi} 
                                                      \left( \hat{D}_\nu\hat{\phi}\right)^* 
+  \frac{1}{2}  \hat{\phi}^2 - \frac{1}{4}  \hat{\phi}^4,  \nonumber \\
                                                \label{eqn:DimensionlessBECLagrangianAppendix}
\end{eqnarray}
where
\begin{eqnarray}
\hat{D}_\mu\hat{\phi} &=& \left( \frac{1}{\kappa}\hat{\partial}_\mu  - i \hat{A}_\mu\right)\hat{\phi}, {\rm \ \ \ and} \\
\hat{F}_{\mu\nu} &=& \hat{\partial}_\mu \hat{A}_\nu - \hat{\partial}_\nu \hat{A}_\mu.
\end{eqnarray}
The Lagrangian (\ref{eqn:DimensionlessBECLagrangianAppendix}) is used throughout this paper 
but with the hats removed for clarity.
The work by Shellard and Ruback \cite{SHELLARD1988262} uses the Lagrangian
\begin{eqnarray}
{L_{\rm SR}} &=& \frac{1}{2}\eta^{\mu\nu}
\left( \partial_\mu - i {A}_\mu\right){\phi}
\left( \partial_\nu + i {A}_\nu\right){\phi}^*  \nonumber \\
&&
-\frac{1}{4}{F}^{\mu\nu}{F}_{\mu\nu}  
-\frac{1}{8} \lambda_{\rm SR} \left({\phi}{\phi}^* - 1 \right)^2,
\end{eqnarray}
while the work by Myers et al. \cite{PhysRevD.45.1355} uses the Lagrangian
\begin{eqnarray}
{L_{\rm MRS}} &=& \eta^{\mu\nu}
\left( \partial_\mu - i e{A}_\mu\right){\phi}
\left( \partial_\nu + i e{A}_\nu\right){\phi}^*  \nonumber \\
&&
-\frac{1}{4}{F}^{\mu\nu}{F}_{\mu\nu}  
-\frac{1}{4} \lambda_{\rm MRS} \left({\phi}^*{\phi} - \sigma^2 \right)^2\!. \ \ 
\end{eqnarray}
These two Lagrangians are physically identical to (\ref{eqn:DimensionlessBECLagrangianAppendix}), 
and they can all be compared to one another with the
conversions in Table \ref{table:LagConversions}.
%
%
\begin{table}[h]
\begin{tabular}{ ccc }
\hline
\hline
$\hat{L}$ 	& $L_{\rm SR}$ 	& $L_{\rm MRS}$  \\
\hline
$\kappa \hat{A}_\mu$ 	& $A_\mu$ & $\sqrt{2}\sigma A_\mu$  \\
$\hat{\phi}$		& ${\phi}$	& $\sigma\phi $\\
$g_{\mu\nu}$ 	& -$\eta_{\mu\nu}$ & -$\eta_{\mu\nu}$ \\
$\hat{\partial}_\mu$	& $\partial_\mu$	& $\sqrt{2}e \sigma {\partial}_\mu$ \\
$2\kappa^2$	& $\lambda_{\rm SR}$ & $\lambda_{\rm MRS}/\!\!\left(2e^2\right)$ \\
\hline 
\hspace{12mm} & \hspace{12mm} & \hspace{12mm} \\
\vspace{-7mm}
\end{tabular}
\caption{
Model parameters for the Ginzburg-Landau model used in this work ($\hat{L}$) and how it compares to the
Abelian-Higgs models of \cite{SHELLARD1988262,PhysRevD.45.1355}. 
Conversion from one model to another is performed by substituting values from one column with their respective values
from another column.
\label{table:LagConversions}
}
\end{table}
A quantity that is commonly discussed is the Ginzburg-Landau (or Abelian-Higgs) coupling parameter
for noninteracting vortices that separates attractive (Type I) vortices from repulsive 
(Type II) vortices.   Table \ref{table:LagConversions} (bottom row) shows that $\kappa = 1/\sqrt{2}$ corresponds
to $\lambda_{\rm SR} = 1$ and $\lambda_{\rm MRS} = 2e^2$ (where the convention $e=1$ is usually used).

\section{Computational Methods \label{app:ComputationalMethods}}

This appendix briefly discusses the numerical methods used in this paper.
All vortex collisions were performed using a second-order implicit Crank-Nicolson 
finite difference scheme coded in C++ and CUDA and were run on a single NVIDIA 
GeForce RTX 2060 Super graphics processing unit (GPU) with 8 GB of RAM.
This GPU platform was ideally suited for these calculations and the grid sizes used;
all of the  grid functions could be stored directly on the GPU device simultaneously 
(roughly 60\% memory utilization for the largest, $\kappa=80$, grids), 
resulting in minimal inefficiencies from moving data over the PCI-e bus.

Given the large dynamic range of length scales, where the magnetic fields change on the order of $\Lambda\approx 1$ 
while the scalar field changes (at a vortex) on the order of the coherence length ($\xi=1/\kappa$), it is imperative that one
use fine enough grid resolution to properly resolve the salient features of the system.
This was done by maintaining a ratio of $\xi/dx > 10$ for all simulations, which led to simulations on  large grids 
(Table \ref{table:SimulationParameters}). 

A common  method used to determine solution quality is to monitor how well energy is conserved in the simulation over time.  
This is done by integrating the energy density in the spatial grid at any time and adding that to 
the integrated momentum flux along  the boundary 
from the beginning of the simulation to the time of interest.
The energy density of the system is given by the sum of the  energy densities,
\begin{eqnarray}
\rho^{kin} 
&=&  
\frac{1}{2\kappa^2}\left[  \Pi_1^2 + \Pi_2^2  + \Phi_{1x}^2 + \Phi_{1y}^2 + \Phi_{2x}^2 + \Phi_{2y}^2 \right], \\
\rho^{\rm cur}	
&=&  
\frac{1}{\kappa}\left[ 
{A}_t \left( \phi_2\Pi_1 - \phi_1\Pi_2 \right) + 
{A}_x \left( \phi_2\Phi_{1x} - \phi_1\Phi_{2x} \right) \right. \nonumber\\
&& \left. + 
{A}_y \left( \phi_2\Phi_{1y} - \phi_1\Phi_{2y} \right)
\right], \\
\rho^{EM} &=& \frac{1}{2}\left[ 
{E}_x^2  + {E}_y^2  + {E}_z^2 
+{B}_x^2  + {B}_y^2  + {B}_z^2 
	\right],   \\
\rho^{mass} &=&  \frac{1}{2}\left( \phi_1^2 + \phi_2^2\right)\left[
{A}_t^2 + {A}_x^2 + {A}_y^2 + {A}_z^2 
\right], {\rm \ \ \ and} \\
\rho^{V} 	
&=& -\frac{1}{2} {\phi}^2 +  \frac{1}{4}{\phi}^4  + \rho^{V}_0, 
\end{eqnarray}
where $\rho^{V}_0$ is set such that $\rho^{V}=0$  when $\phi$ is at the expectation value of the
spontaneously broken symmetry.
The momentum densities of the system are given by
\begin{eqnarray}
{j}_x^{\rm kin} 	&=& -\frac{1}{\kappa^2}\left( {\Phi}_{1x}{\Pi}_1 + {\Phi}_{2x}{\Pi}_2\right),  \\
{j}_y^{\rm kin} 	&=& -\frac{1}{\kappa^2}\left( {\Phi}_{1y}{\Pi}_1 + {\Phi}_{2y}{\Pi}_2\right),  \\
{j}_x^{\rm cur} 	&=& -\frac{1}{\kappa}\left[   
{A}_t \left( {\phi}_2 {\Phi}_{1x} - {\phi}_1{\Phi}_{2x}  \right)  \right. \nonumber \\
&& \left.  +
{A}_x \left( {\phi}_2 {\Pi}_{1} - {\phi}_1{\Pi}_{2}  \right)
\right], \\
{j}_y^{\rm cur} 	&=&  -\frac{1}{\kappa}\left[   
{A}_t \left( {\phi}_2 {\Phi}_{1y} - {\phi}_1{\Phi}_{2y}  \right)  \right. \nonumber \\
&& \left.  +
{A}_y \left( {\phi}_2 {\Pi}_{1} - {\phi}_1{\Pi}_{2}  \right)
\right], \\
{j}_x^{\rm EM} 	&=& {E}_y{B}_z - {E}_z{B}_y, \\
{j}_y^{\rm EM} 	&=& {E}_z{B}_x - {E}_x{B}_z, \\
{j}_x^{\rm mass} 	&=& -\left( {\phi}_1^2 + {\phi}_2^2\right) {A}_x {A}_t,   {\rm \ \ \  and}    \\
{j}_y^{\rm mass} 	&=& -\left( {\phi}_1^2 + {\phi}_2^2\right) {A}_y {A}_t.
\end{eqnarray}
%
\begin{figure}[t]
\includegraphics[width=0.9\linewidth,height=0.91\linewidth]{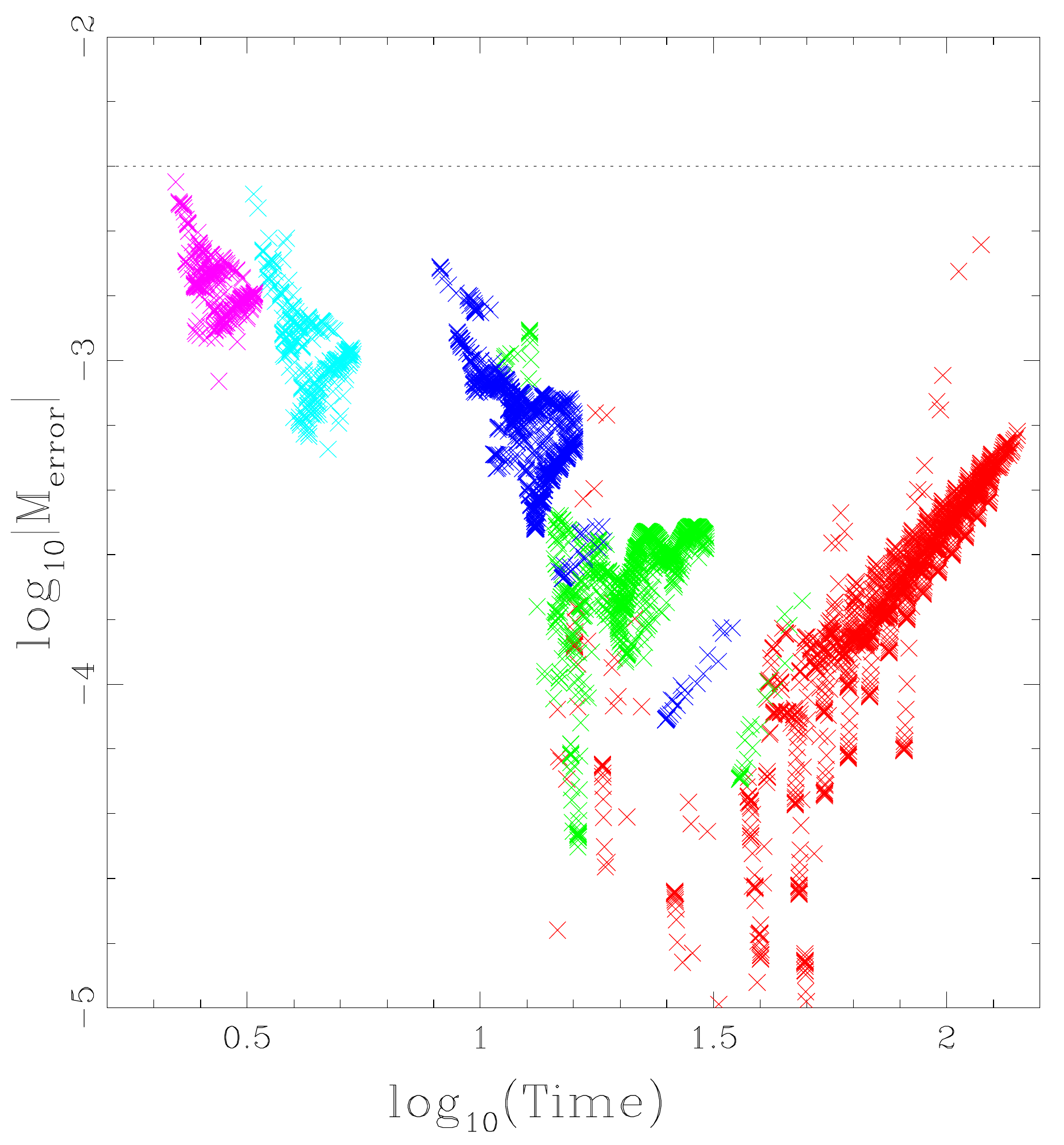}%
\caption{ Plots of final $M_{\rm error}$ all 10,867 simulations from Figs. 
\ref{fig:DeflectionLog}, \ref{fig:CS_4_10_A}, and \ref{fig:CS_40_80}. 
Simulations with $\kappa = 1, 4, 10, 40,\ {\rm and }\ 80$ are plotted in red, green, blue, cyan, and magenta, 
respectively. 
Most simulations had $M_{\rm error}$ on the order of $10^{-3}$, while all simulations had 
$M_{\rm error} < 4\times10^{-3}$ (dotted line).
\label{fig:EnergyConservation}}
\end{figure}
%
Figure \ref{fig:EnergyConservation} shows the final mass error, 
\begin{equation}
M_{\rm error} =  \frac{M_{\rm final} + M_{\rm radiated} - M_{\rm initial}}{M_{\rm initial}},
\end{equation}
where $M_{\rm final}$ is the integrated energy density at the end of the simulation, 
$M_{\rm initial}$ is the integrated energy density at $t=0$ , 
and $M_{\rm radiated}$ is the integrated outgoing momentum flux over the outer boundary of the computational
domain  over the entire time of the simulation.
Most of the simulations conducted had final $M_{\rm error}$ between $10^{-4}$ and $10^{-3}$. 
While the ratio $\xi/dx > 10$ was maintained for all simulations, 
the increase in  $v_0^*$ as $\kappa$ increases leads to 
Lorentz contraction of the vortex in the $x$ direction by a factor of $\gamma_0^*$ that reduces the
effective number of grid points per vortex in the direction of motion, 
leading to lower effective resolution and higher final $M_{\rm error}$ values for larger $\kappa$.
Despite this reduction in resolution, the simulations conducted here have typical mass errors that are still 1, 
if not 2, orders of magnitude lower than previous work.

\vspace{3cm}

\bibliography{VortexPaper}

\end{document}